\documentclass[letterpaper,journal]{IEEEtran}

\usepackage{algorithmic}
\usepackage{algorithm}
\usepackage{array}
\usepackage[caption=false,font=normalsize,labelfont=sf,textfont=sf]{subfig}
\usepackage{textcomp}
\usepackage{stfloats}
\usepackage{url}
\usepackage{verbatim}
\usepackage{graphicx}
\usepackage{cite}
\usepackage{stfloats}

\usepackage{comment}
\usepackage{amsmath,amssymb,amsfonts}
\usepackage{graphicx}
\usepackage[export]{adjustbox}
\usepackage{multirow}
\usepackage{makecell}
\usepackage{bm}
\usepackage{booktabs}
\usepackage[dvipsnames]{xcolor}
\usepackage{pgf}

\usepackage{colortbl}
\usepackage{xcolor}

\begin{document}

\title{FoveaSPAD: Exploiting Depth Priors for Adaptive and Efficient Single-Photon 3D Imaging}
% Make Title

\author{
  Justin Folden, Atul Ingle$^*$, and Sanjeev J. Koppal$^*$%
\thanks{Manuscript received XXXX YYY, 2024; revised XXXX YYY, 2024.}
   \thanks{($^*$ equal contribution)}
  \thanks{J. Folden and S. Koppal are with the Department of Electrical and Computer Engineering, University of Florida, Gainsville, FL. Email: jfolden@ufl.edu.}% <-this % stops a space
\thanks{A. Ingle is with the Department of Computer Science, Portland State University, Portland, OR.}% <-this % stops a space
\thanks{This paper has supplementary downloadable material available at http://ieeexplore.ieee.org., provided by the author. The material includes an mp4 video, as well as a supplemental pdf. This material is 45MB in size.}
}% <-this % stops a space

\maketitle

\begin{abstract}
Fast, efficient, and accurate depth-sensing is important for safety-critical applications such as autonomous vehicles. 
Direct time-of-flight LiDAR has the potential to fulfill these demands, thanks to its ability to provide high-precision depth measurements at long standoff distances.
While conventional LiDAR relies on avalanche photodiodes (APDs), single-photon avalanche diodes (SPADs) are an emerging image-sensing technology that offer many advantages such as extreme sensitivity and time resolution.
In this paper, we remove the key challenges to widespread adoption of SPAD-based LiDARs: their susceptibility to ambient light and the large amount of raw photon data that must be processed to obtain in-pixel depth estimates.
We propose new algorithms and sensing policies that improve signal-to-noise ratio (SNR) and increase computing and memory efficiency for SPAD-based LiDARs. 
During capture, we use external signals to \emph{foveate}, i.e., guide how the SPAD system estimates scene depths. 
This foveated approach allows our method to ``zoom into'' the signal of interest, reducing the amount of raw photon data that needs to be stored and transferred from the SPAD sensor, while also improving resilience to ambient light. We show results both in simulation and also with real hardware emulation, with specific implementations achieving a 1548-fold reduction in memory usage,  and our algorithms can be applied to newly available and future SPAD arrays. 

% However, conventional LiDARs that rely on avalanche photodiodes (APDs) suffer from high cost, limited spatial resolution, and low framerates. Single-photon avalanche diodes (SPADs) are an emerging image-sensing technology for solid-state direct time-of-flight depth sensing. Thanks to their extreme sensitivity, and high pixel resolutions approaching kilo-to-megapixels, SPADs have positioned themselves as a promising alternative to traditional APD sensors. A key challenge to widespread adoption of SPAD-based LiDARs is their susceptibility to ambient light and the large amount of raw photon data that must be processed to obtain in-pixel depth estimates. 

% Here we propose a novel foveated SPAD LiDAR that is externally guided through the use of a single RGB camera and a state-of-the-art monocular depth estimation network. This foveated approach allows our method to ``zoom into'' the signal of interest, reducing the amount of raw photon data that needs to be stored and transferred from the SPAD sensor, while also improving resilience to ambient light.
\end{abstract}


% The first section title should be wrapped inside a \IEEEraisesectionheading as follows.
\section{Introduction}\label{sec:introduction}
% \begin{itemize}
% \item Big picture: resource aware 3D sensing of scene information.
% 
% \item Inspiration from biology, explain what foveation means. Biological vision systems foveate in depth using stereo cues to adjust the pupil aperture/depth of field. [Q: Do bats and dolphins foveate when echolocating?]
% 
% \item Wouldn't it be cool if we can foveate in the 3rd dimension too! (Put some ``marketing'' stuff here --- which applications could benefit if we could foveate in 3D?)
% 
% \item Foveation with SPAD studied before under the garb of partial histograms, and more recently as zooming and sliding and range gating.
% 
% \item SPAD naturally suited for depth foveation due to their high speed and gating capabilities at the granularity of single photons, at sub nanosecond time scales.
% 
% \item Contributions (not in any specific order):
% \begin{itemize}
% \item We contextualize this scattered literature and present a unified formalism in terms of "foveation in depth".
% \item We present techniques that optimize memory and resolution.
% \item We explore the question of how to optimally foveate in depth --- monocular estimates, exploiting optical flow/motion.
% \end{itemize}
% \end{itemize}
% 
% Other thoughts:
% Can we foveate both spatially and in depth? Imagine reorganizing smaller / larger macro pixels with fat/narrow bins adaptively.

\IEEEPARstart{B}{iological} vision systems have the remarkable ability to \emph{foveate} --- i.e. redistribute cognitive resources towards ``salient'' features or objects in a scene, depending on context.
Unfortunately, most conventional cameras and computer vision systems today capture scene information in a non-adaptive fashion, spending power and bandwidth on sensing scene components that may not help the overall imaging task. In fact, the current framework for deep learning-based systems assumes uniform sampling of the scene and overcomes these limitations through data-driven pipelines that focus on interesting regions of the scene \cite{vaswani2017attention} \cite{selvaraju2017grad} in the input RGB images.

\begin{figure*}
 \includegraphics[width=\textwidth]{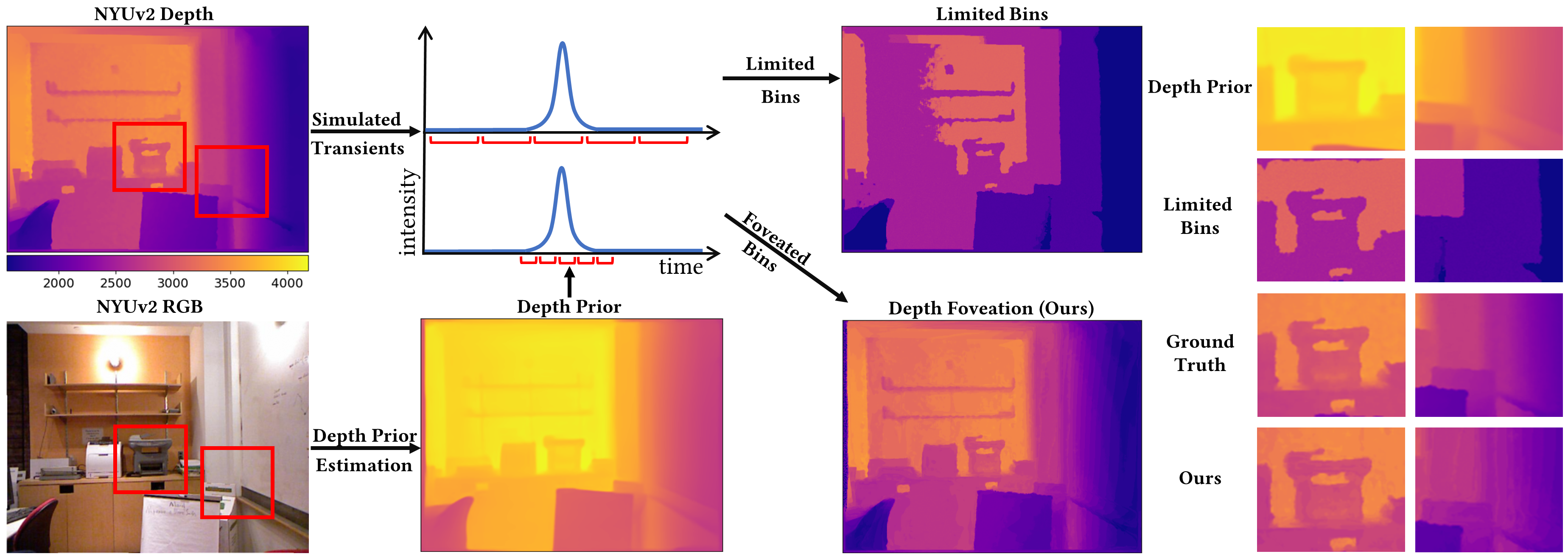}
 \caption{\textbf{Depth Prior Driven SPAD Depth Foveation:} SPAD sensors suffer from a data bottleneck, since thousands of histogram bins are used to generate depth as shown in the top left. If fewer bins are used, this reduces depth resolution, as shown in the limited bins depth result. Our idea is to use additional information, such as a color image (Sec. \ref{sec:4monofovea}, \ref{sec:hardware}) or optical flow (Sec. \ref{sec:6opflow}), to foveate the SPAD bins. Therefore, for the same memory cost we can place the bins near where the histogram peak should be, results in accurate depth, as shown in the depth foveation result. The insets show that our method achieves the accuracy and resolution of ground truth, with fewer bins. They also show that the depth prior, in this case monocular estimation, by itself cannot provide the correct depth, and foveation is required.}
 %\Description{Our technique for saving memory, and improving depth resolution for SPAD cameras.}
 \label{fig:teaser}
\end{figure*}

While this inefficient but popular framework for conventional RGB sensors may be difficult to change, our proposed method, called FoveaSPAD, can impact the next wave of single-photon avalanche diode (SPAD) sensor technology. SPADs can capture scene information at the granularity of individual photons, at timescales as small as 10's of picoseconds. Recent advances in CMOS-compatible SPAD pixel designs has enabled real-time in-pixel processing of these photon timestamp streams. Thus, SPADs are a natural candidate for designing efficient depth cameras --- individual pixels can be reprogrammed on-the-fly to adaptively accept or reject a spatio-temporal subset of the photon stream.

Our FoveaSPAD algorithms enable capturing scene information at higher granularity in regions that are most relevant to a downstream vision task. \emph{In this sense, we generalize the term ``foveation'' in the context of adaptive SPAD spatio-temporal sampling to allow both depth and memory efficiencies.}
For robots, remote sensor nodes, and other resource-constrained systems, foveation for SPAD sensors can allow accurate depth sensing under constraints on power and bandwidth (see Fig. \ref{fig:teaser}). 

The raw data captured by an array of SPAD pixels can be thought of as a spatio-temporal photon stream. Each photon detection is represented as an $(x,y,t)$ coordinate, where the $x-y$ coordinates denote the pixel location and the $t$ coordinate denotes the photon detection timestamp. Each SPAD pixel captures the round trip time of a laser pulse to and from a given scene point, constructing a photon timing \emph{histogram} which records the number of photons captured at various time delays with respect to the time the laser pulse was transmitted. Each pixel must construct one such histogram, typically with 1000's of bins, which causes a severe data bottleneck for today's SPAD cameras. To illustrate the severity of the bottleneck, consider a 1-megapixel SPAD array with a 1000-bin histogram per pixel, storing 1 byte per bin. At 30 frames per second, this setup generates a staggering 30 GB of data every second

Our algorithms foveate across the spatio-temporal histogram space to efficiently recover the peak, providing the time-delay $t$ for depth computation. We adaptively capture subranges to locate laser photons, rejecting ambient photons.
Note that our proposed algorithms are not exhaustive; rather, we aim to define a class of algorithms that rely on a depth prior. In this work, we propose three methods for acquiring priors, though many other methods exist, such as depth from stereo, depth from defocus \cite{Xiong1993defocus}, or non-vision-based methods such as sonar. Each method has trade-offs, and it is up to the user to determine which method best suits their use case. Our contributions in this work are as follows:

%This has enabled several new imaging capabilities such as light-in-flight (transient) imaging \cite{}, non-line-of-sight imaging \cite{}, fluorescence lifetime microscopy \cite{}, and high resolution 3D imaging LiDARs.

% subsection set before our contributions 
% Subsection: Our proposed algorithms are not comprehensive
% We propose 3, but there can be many more.
    % we define a class of algorithms
    % could be a nonvisual sensor, sonar, depth from defocus, etc
    % Each has tradeoffs, up the the user
    % We can use _stereo_ as a prior
    
% Here's the math that're independant of the prior.
    %others might find useful

\begin{itemize}
  \item We present a theoretical model for expected gains (in terms of increased signal-to-noise-ratio and depth resolution) from foveation with SPADs.  
  \item We explore the question of how to foveate in space and time at a single time instant by leveraging monocular depth estimates, which can come either from the SPAD-generated image or a cheap, external color camera. We propose different flavors of practical FoveaSPAD designs that optimize for memory/bandwidth and depth resolution.
  \item For images of moving scenes, we demonstrate how to use optical flow cues to direct SPAD foveation. 
%  \item We provide a worst-case analysis of the limits of foveation with ambient light in SPAD sensors. 
  \item We show results both in simulation and using recently available real SPAD datasets.
\end{itemize}

% \smallskip
% \noindent\textbf{Limitations and Scope:} 

\subsection{Hardware Emulation}

Time-correlated single photon counting is the technique that enables SPAD cameras to build histograms and control binning on-sensor. Our work is limited to simulation experiments and hardware emulation of existing SPAD LiDAR data. Hardware emulation refers to leveraging real-world data captured using a single SPAD or line arrays \cite{lindell2018single,gutierrez2022compressive}, which we then use to emulate the performance of larger SPAD arrays. While SPAD sensor arrays with native support for foveation are not yet available commercially, we believe our proposed techniques could be implemented at the pixel or camera level. This is supported by recent proofs-of-concept in kilopixel-resolution reconfigurable SPAD arrays with in-pixel timestamping, gating, and histogramming capabilities \cite{hutchings2019reconfigurable,zhang201830}. We anticipate that this work will inspire future hardware designs, leading to more efficient and versatile SPAD sensor arrays.

\subsection{Scope: Simulation and Emulations}
In this work, we anticipate future hardware advancements that will enhance SPAD-based depth sensing. Our simulations and emulations are intended to project the performance of emerging SPAD sensor technologies, focusing on adaptive and efficient bin sampling to mitigate memory bottlenecks with minimal loss of accuracy, which is particularly advantageous for flash-based SPAD LiDARs systems. Potential future implementations could feature a shared ``macropixel'' architecture and a dynamic gating system, allowing pixel groups to adjust to appropriate gating signals in real time. We explore these ideas further in Sec. \ref{sec:limit} and present a speculative ``macropixel'' array design in Figure \ref{fig:pixeldesign}, which includes a variable-resolution TDC—a key component for one of the proposed methods. These simulations play a critical role in validating our algorithms and highlighting their potential impact on future sensor designs, even in the absence of current hardware.

\section{Related Work}

Our research takes inspiration from biology, since many animals have a region of high spatial acuity, i.e. the fovea, which they scan over the scene. In this sense, we are allied with foveated imaging research in computer vision and computational photography, and we now outline these related efforts:

\smallskip
\noindent\textbf{Efficiency in Single-photon 3D Cameras:} The data bottleneck issue in SPADs due to high-resolution sampling in histograms is well-known. Research that attempts to mitigate this issue include novel statistical representations \cite{heide2018sub} as well as compressive histograms \cite{gyongy2020high, gariepy2016picosecond, gutierrez2023learned} that use a small number of bins at maximum resolution to recover the entire scene. 
In contrast, our approach works and scales easily with a large number of SPAD pixels. Other efforts include partial histogram methods such as using sliding windows for sub-range gating has been investigated \cite{ren2018high, dutton2015spad, erdogan2019cmos, dutton2016single} which have linear efficiency and two-stage coarse-to-fine resolution scaling \cite{zhang2021240} which provide logarithmic efficiency. Our method uses context from cues such as optical flow to provide $\approx O(1)$ near-constant time efficiency. Finally, other work has used external sensors for guided upsampling or upscaling, \cite{lindell2018single, taneski2023guided}, but these are post-capture processes. In contrast, we perform foveation during capture and this gives us SNR and compute efficiencies that we have theoretically analyzed.
A complementary approach to foveation is to use adaptive ``equi-depth'' histogramming approach for the signal peak \cite{ingle2023count}. Our approach is also complementary to adaptive gating approaches for SPAD LiDARs \cite{po2022adaptive}, with adaptive gating and exposure techniques working with or without a prior. 

\smallskip
\noindent\textbf{Foveated Depth Sensors: } Our work is related to post-capture methods for upsampling and superresolution shown on data from many modes, such as depth images, color photographs etc. ~\cite{battrawy2019lidar,chen2018estimating,mal2018sparse,lu2015sparse,uhrig2017sparsity} and many of these have blended deep learning algorithms into the process of deciding where to sample
\cite{riegler2016atgv,hui16msgnet, uhrig2017sparsity, van2019sparse,gruber2019gated2depth, tilmon2021saccadecam}. In fact, some of these algorithms are mature enough that commercial depth and LIDAR sensors allow post-capture foveation of the 3D point cloud through, for example, LIDAR-RGB fusion. In contrast, FoveaSPAD adapts during capture, and the efficiencies can impact small autonomous systems with power constraints. Directionally controlled LIDAR systems foveate spatially \cite{yamamoto2018efficient, tasneem2018dirrectionally, adaptivelidarstanford, pittaluga2020towards}. These results complement our work on temporal foveation of SPAD sensors, including spatio-temporal foveation results (Sec. \ref{sec:5ST}).

\smallskip
\noindent\textbf{Foveation in Display Graphics: } Foveation is an important research topic in computer graphics, where data displayed to a viewer on AR/VR glasses, for example, is rendered in a way that reduces bandwidth \cite{guenter2012foveated}. Most of the work in this area does not focus on data capture but only on data visualization post-capture \cite{albert2017latency,tursun2019luminance}. Foveated light-field optics have been proposed \cite{huang2015light} and these can be integrated with algorithms that foveate which portions of the scene to render at high resolution to reduce rendering resource consumption. Algorithms include perceptually guided foveation \cite{sun2017perceptually,patney2016perceptually} and hardware-optimized rendering \cite{meng20203d}. Unlike our depth sensor, these use passive displays and cameras to optimize bandwidth, storage, and compute.

\smallskip
% Redundant to Efficency. Add the extra sources from R3 and make this part about those.
\noindent\textbf{SPAD Histogram Techniques:} Various techniques have been recently proposed to reduce the memory and bandwidth required to capture high-resolution photon timing histograms.
Compressive histogramming techniques rely on a lower-dimensional linear projection of the high resolution histogram\cite{gutierrez2022compressive, gutierrez2023learned} and estimating scene distances directly from the compressed representation. 
Algorithms that rely on ``sketching'' \cite{sheehan2021sketching} attempt to directly estimate a parametric form of the true underlying waveform.
These compressive acquisition approaches can be combined with foveation techniques developed here to further reduce the bandwidth required to store histograms.
Differential capture methods \cite{zhang2022firstarrival,white2022differentialspad} can provide large reduction in bandwidth, but unlike foveation-based techniques, differential capture methods require additional post-processing to recover absolute scene depths.
Recently, photon processing techniques that bypass the need for constructing a histogram have also been proposed, but they only work in the case of a single strong peak \cite{totini2023histogramlesslidar}.
Sun et al.'s optical coding and super-resolution techniques leverage a phase plate and deep learning to achieve super-resolved images with minimal photon counts, further optimizing SPAD-based imaging \cite{sun2020superresspad}. Such optical techniques can work synergistically with our foveated capture approach, collectively reducing data transfer and computational demands.

\begin{comment}
- Repeat this is a simulation and cite code and paper. 
- Number of photons for all experiments is the same, but memory and depth resolution from foveation changes. No decrease in time, 
- Every groundtruth depth image in this paper has been created from a full histogram transient and is not simple copied depth from the dataset
- color images could be obtained in a colocated manner, since color SPADs are now available CITE
\end{comment}
\section{Imaging Model and the Foveation Advantage}
\label{sec:3advantage}
In this section, we present the imaging model and the concept of foveation, specifically focusing on how foveation can enhance the efficiency and effectiveness of (SPAD) LiDAR systems. We will delve into the specifics of how the imaging model is constructed, including assumptions about the behavior of laser pulses and photon detection, and how these factors influence the design and performance of SPAD sensors. Furthermore, the impact of ambient light on signal-to-noise and signal-to-background ratios will be examined, demonstrating how foveation can mitigate these effects. The theoretical foundations laid out in this section will serve as the basis for the foveation techniques proposed in the subsequent sections, where we will develop and analyze algorithms to optimize the selection of foveated bins in SPAD imaging.

\subsection{Foveation and Scene Priors}
We propose two methods of foveation, specifically memory foveation and depth foveation, are designed to optimize the efficiency of SPAD LiDAR systems by leveraging a priori knowledge about the scene's depth. Both methods require adaptive per-pixel gating, for which the hardware has yet to be developed.

Memory foveation focuses on reducing the amount of data that needs to be stored and processed by concentrating on a subset of histogram bins where the depth information is most likely to reside. Depth foveation, on the other hand, aims to improve depth resolution by reallocating histogram bins into a smaller, more focused region around the expected depth. The strategies proposed are fundamentally dependent on the accuracy and reliability of the scene depth prior, which guide the allocation of sensor resources.

Depth priors may be derived from any variety of means, including coarse initial scans, external sensors, or deep learning models. In this paper, we explore a few options, namely monocular estimation in Sect. \ref{sec:4monofovea}, optical flow warping in Sect. \ref{sec:6opflow},  and coarse initial scans in Sect. \ref{sec:hardware}. The quality of the prior directly impacts the success of foveation, with inaccurate priors potentially misallocating memory resources into incorrect regions. This dependence implies a trade-space between depth prior accuracy, and the amount of resources foveation stands to reduce. Exploring this trade-space is out of the scope of this paper, rather, we focus on using priors that are prone to error or are otherwise lower quality.

In the following subsections, we will define the image formation model, detailing the assumptions and mechanics of photon detection. We will then explore the effects of ambient light on SPAD histogram formation and discuss how the proposed foveation techniques provide an advantage.
{

\footnotesize

\begin{table}
\caption{Mathematical symbols used in this paper to study the foveated SPAD imaging model.\label{tab:symbols}}
\centering
\begin{tabular}[t!]{ p{2cm} p{6cm} } 
 \toprule
 Symbol & Meaning  \\ 
 \midrule
 $N$ & Number of bins across full histogram  \\ 
 $M$ & Number of bins across foveated histogram  \\
 $i$ & Bin location of corresponding to true scene depth \\
 $Z$ & Working volume of the sensor \\
 $T$ & Temporal volume calculated from Z and speed of light \\
 SNR & Signal-to-noise ratio \\
 SBR & Signal-to-background ratio \\ 
 $C$ & Number of cycles to create histogram \\
 $\Phi_\text{sig}$ & Mean number of signal photons received per bin \\
 $\Phi_\text{bkg}$ & Mean number of background photons received per bin \\
 $p_\text{gt}$ & Probability that a detected photon originated from the laser \\
 $p_\text{multipath}$ & Probability that a detected photon experienced multipath bounces  \\
 $p_\text{floor}$ & Probability of a low noise floor \\
 $S$ & Number of pixels in the camera \\
 \bottomrule 
\end{tabular} 

\end{table}

}

\subsection{Image Formation Model}
We assume that each pixel in the SPAD sensor array is co-located with a pulsed laser illumination source with a Gaussian pulse shape.
Assuming no multi-path or sub-surface scattering effects, the photon flux incident on each pixel consists of a superposition of laser photons (that arrive in a short time window corresponding to the round-trip time-of-flight to and from the scene point) and background photons due to ambient light (that arrive uniformly randomly distributed throughout the capture duration).
The laser repetition period ($T$) determines the maximum depth range of the SPAD LiDAR.
We assume that this period is discretized into $N$ bins ($N$ is often on the order of 1000's of bins in conventional SPAD cameras).
The number of photons captured by the SPAD pixel in the $n^\text{th}$ bin ($1\leq n \leq N$) is Poisson distributed with a mean of $\Phi_\text{sig} \mathbf{1}(n=i) + \Phi_\text{bkg}$ where $i$ is the bin location corresponding to the true scene depth.
Various sources of noise such as dark counts and afterpulsing are assumed to be absorbed in the $\Phi_\text{bkg}$ term.
A complete histogram captured by this SPAD pixel over $C$ laser cycles is given by a Poisson random vector with mean $ C\Phi_\text{sig} \mathbf{1}(n=i) + C\Phi_\text{bkg}$ for $1 \leq i \leq N.$

The simplified imaging model assumes all laser photons arrive in a single bin $i$.
In practice, the laser pulse spans several bins ``smearing'' the signal photons over more than one bin.
The laser peak is often modeled as a Gaussian shaped pulse; we use a 1 nanosecond full width at half maximum (FWHM) in our simulation results. Since the peak can span more than one histogram bin location, the defined Gaussian pulse may be used to estimate depth through match filtering.
It is also possible to obtain a pseudo-intensity image by aggregating photon counts across histograms for each pixel which can be used in lieu of a co-located RGB or monochrome camera image for monocular depth cues.

\subsection{Effects of Ambient Light}
The integration time taken for all experiments is consistent. In this scenario, we show how  foveation saves memory or improves depth resolution, and how the signal-to-noise ratio changes depending on ambient light, bin width, and the number of laser cycles or exposure time.
%Atul: Not sure what "low sensor noise floor means --- is this refering to, say, a spurious "dark count" event being detected in a particular bin?

Consider a SPAD pixel imaging a scene point illuminated by a pulsed laser. Initially, let us assume there are no multi-bounce effects and no ambient light, although we address these issues later on. 

Photon detections from the SPAD pixel generate a histogram of arrival times. A conventional approach would use all $N$ bins across the full histogram, whereas we propose methods to foveate attention onto a subset $M \le N$ of these bins, where $M$ is a window or gate with a user defined width (number of bins). Therefore, it is not surprising that, in the SNR analysis of our system, the ratio $\frac{M}{N}$ appears since this represents the advantage due to foveation. 

In the analysis below, we will not make any assumption as to how the foveated bins $M$ were obtained and instead just characterize the advantage of these, given that the desired histogram peak is captured by these bins. The analysis is not specific to any one method of acquiring a depth prior. In Sections \ref{sec:4monofovea}, \ref{sec:5ST}, \ref{sec:6opflow}, and \ref{sec:hardware} we propose algorithms to drive the selection of the foveated bins $M$ and in Sect. \ref{sec:limit} we provide a worst case analysis for whether the foveated $M$ bins capture the histogram peak or not.
% ?? The anal% 
\subsubsection{Low Ambient Light (No Pileup)}
\label{subsec:Nopileup}

Now consider the conventional imaging case, where the SPAD sensor detects time-of-arrival of photons and accumulates into a photon timing histogram to find the time that corresponds to the true depth of the scene point.

We assume that the histogram has a full scale range of $T$ seconds which is related to the maximum unambiguous depth range $Z$ as $T = \frac{2Z}{c}$ where $c$ is the speed of light.
Consider $N$ histogram bins that are uniformly distributed across the full scale range $T$.
The width of each bin is $\frac{T}{N}$.
Since narrower bins produce fewer photons, the SNR for each bin is proportional to the width of that time bin: 
\begin{equation}
   \text{SNR} \propto C \ \sqrt{\frac{T}{N}},   
\end{equation}
where $C$ denotes the number of laser cycles (i.e., the total exposure time) that was used to capture the histogram.

%The signal-to-background ratio (SBR) is defined \cite{gupta2019asynchronous} as  the ratio of the total number of signal photons to the total number of background photons received over each laser cycle, $\mathsf{SBR}=\frac{\Phi_{sig}}{N \Phi_{bkg}}$.
%Here $\Phi_\text{sig}$ is the mean number of signal (i.e. laser) photons received per laser cycle and $\Phi_\text{bkg}$ is the mean number of background (e.g. ambient) photons received per bin.
%The SBR decreases as the number of bins increases:
%\begin{equation}
%   SBR \propto \frac{1}{N}. \leftarrow \text{not correct}
%\end{equation}

We now consider two types of foveation. In \emph{memory foveation}, only a limited number of bytes in memory can be dedicated to the task of finding the histogram peak, and therefore placing these at the peak is most efficient. In \emph{depth foveation}, memory allocation remains fixed but is concentrated in the foveated region, bringing the bins closer together near the histogram peak, thereby improving depth resolution.

\smallskip
\noindent
\textbf{Memory foveation:} In memory foveation, we identify $M$ bins $M \ll N$ where the true depth exists. The width of the bins remains the same $\frac{T}{N}$, and therefore the SNR is also identical to the conventional case: 

\begin{equation}
   \text{SNR} \propto \sqrt{\frac{M \frac{T}{N}}{M}} \propto\sqrt{\frac{T}{N}} 
\end{equation}

%However, the SBR is now higher by a factor $N/M$.
%Since we search over only a few number of bins $M$ the total number of background photons is now $M\Phi_\text{bkg}$ instead of $N\Phi_\text{bkg}$, but the total number of signal photons stays unchanged at $\Phi_\text{sig}$.
%\begin{equation}
%   SBR \propto \sqrt{\frac{1}{M}}. \leftarrow \text{not correct}
%\end{equation}
\smallskip\noindent
\textbf{Depth foveation:} In depth foveation, we concentrate the $N$ bins that would have been distributed over the entire depth range, into a small region. The region is the same region used in memory foveation, and is given by multiplying the number of memory foveation bins $M$ with the original bin width to give $M \ \frac{T}{N}$. This region is divided into $N$ bins, and therefore the new bin width is $\frac{MT}{N^2}$. As before, the SNR is proportional to the bin width, and therefore much lower, 

\begin{equation}
   \text{SNR} \propto C \ \sqrt{\frac{MT}{N^2}} = \sqrt{\frac{M}{N}\frac{T}{N}}
\end{equation}

Therefore, we have improved depth resolution but at the cost of SNR. To increase the SNR of the foveated depth we can increase $C$, the number of cycles the laser pulses through to create the histogram. The new cycle number must be equal to or greater than $\frac{C_\text{new}}{C} \ge \frac{N^2}{M^2}$, then,

\begin{equation}
   \text{SNR}_\text{new} \propto C_\text{new} \ \sqrt{\frac{MT}{N^2}} =  C \ \sqrt{\frac{T}{N}}.
\end{equation}

In summary, memory foveation reduces memory usage with no change in SNR. Depth foveation increases depth resolution but with reduced SNR that can be compensated by more laser photons (i.e. longer exposure). 

Below, in alg. \ref{alg:general}, we define the general algorithm for memory and depth foveation. Note that the algorithms are independent of depth prior, and the spatio-temporal step, which we show in sec. \ref{sec:5ST}, is optional. 

% \algsetup{linenodelimiter=\tiny}
\def\NoNumber#1{{\def\alglinenumber##1{}\State #1}\addtocounter{ALG@line}{-1}}
\begin{algorithm}
\caption{Memory and Depth Foveation}
\label{alg:general}
\begin{algorithmic}[1]
\REQUIRE Total histogram bins $N$, Temporal Volume $T$, Number of foveated bins $M$, Total histogram bins for depth foveation $N'$
\STATE \textbf{Calculate bin widths}

\hspace{0.5cm} $\Delta t = \frac{T}{N}$, $\Delta t_{depth} = \frac{T}{N'}$
\STATE \textbf{Acquire a depth prior: } 

\hspace{0.5cm} Monocular Sec. \ref{sec:4monofovea}, Optical-Flow Sec. \ref{sec:6opflow}, Low-Resolution Super-Pixel Sampling Sec. \ref{sec:hardware}   

\FOR{$(x, y) \in S$} \label{alg:line:beginFor}
    
    \STATE Utilize the depth prior to find $\hat{d}(x, y)$
    \STATE Center foveation window $M$ around $\hat{d}(x, y)$
    
    \textbf{Memory Foveation:} 
        \STATE \hspace{0.5cm} Capture histogram in the foveated window with bin width $\Delta t$ and $M$ number of bins
        
    \textbf{Depth Foveation:} 
        \STATE \hspace{0.5cm} Capture histogram in the foveated window with bin width $\Delta t_{depth}$ and $N'$ number of bins
\ENDFOR

\STATE \textbf{return} Histogram image $H$
\STATE Decode depth image $D$. $H \rightarrow D$ \label{alg:line:decode}
\vspace{0.5cm}

\textbf{Optional Spatio-Temporal steps:}
\STATE \hspace{0.25cm} \textbf{Quantization Based Sampling} Sec. \ref{sec:5ST}
\STATE \hspace{0.5cm} Quantize depth prior into discrete buckets $B$
\STATE \hspace{0.5cm} Select several pixels in each bucket at random. $S \rightarrow \hat{S}$
\STATE \hspace{0.5cm} Complete steps \ref{alg:line:beginFor}-\ref{alg:line:decode} with $\hat{S}$
\STATE \hspace{0.5cm} Quantize sparse depth map. $ D(B) = \min(D(\hat{S}) \in B)$

\STATE \hspace{0.25cm} \textbf{SuperPixel Based Sampling} Sec. \ref{sec:hardware}
\STATE \hspace{0.5cm} Acquire a pseudo-intensity map through photon counting
\STATE \hspace{0.5cm} Apply the superpixel algorithm to segment the pseud-intensity map
\STATE \hspace{0.5cm} Sample the centroid of each superpixel segment at full histogram resolution. $\hat{d}_\text{SP}$
\STATE \hspace{0.5cm} Complete steps \ref{alg:line:beginFor}-\ref{alg:line:decode} with $S$ and $\hat{d}_\text{SP}$

\end{algorithmic}
\end{algorithm}

%Since the same portion of the volume is sensed in both the depth foveation and memory foveation cases, the SBR is unchanged: 

%\begin{equation}
%   SBR \propto \sqrt{\frac{1}{M}}.
%\end{equation} 

\subsubsection{Strong Ambient Light (Pileup)}
\label{subsec:pileup}

With strong ambient light, we now focus on the signal-to-background ratio (SBR), defined in \cite{gupta2019asynchronous} for SPADs as the ratio of the total number of signal photons to the total number of background photons received over each laser cycle. W.l.o.g, here we note that the SBR is proportional to the probability of receiving signal photons divided by the probability of receiving background photons.

%NOTES: If we want to redo the math here, then the idea is the following. Let the SBR be the ratio of the probability of seeing a ANY photon in the ith bin, with the probability of detecting an ambient photon ANYWHERE (including the ith bin). This second term should be the addition of probabilities of detecting in the 1st bin, plus (not 1st bin)*(2nd bin) + not 1st and 2ndbin)*3rd bin all the way to the k-1th bin. 

%When you foveate, the numerator changes, but the denominator loses terms. In the best case, the denominator loses ALL terms and the numerator just becomes the probability that the ith bin has some photon detected. 

With ambient light, photons from both the laser source and the ambient illumination may be measured by the SPAD. Each time a photon is detected, the SPAD sensor resets creating a pause. It is this pause that creates a binomial model for image capture in SPADs \cite{gupta2019photon,gupta2019asynchronous}.

Therefore, the SBR analysis cannot simply compare the photon bin widths as in the prior section for the full resolution ($N$ bins) and the foveated resolution ($M$ bins). Instead, SBR calculations must include the \emph{probability} of photons from the source vs. the background. 

\smallskip
\noindent \textbf{Conventional scenario:} Let us first consider the SBR in the conventional case, with no foveation. From \cite{gupta2019photon}, using the Poisson model for photon distribution, we can write the probability of a photon from the laser incident on the bin corresponding to the correct depth as $p_\text{laser} = (1 - e^{- \Phi_\text{sig}})$. Correct depth detection will happen even if an ambient photon is detected at the correct depth, so the probability of correct depth detection is $p_\text{correct} = (1 - e^{- (\Phi_\text{sig} + \Phi_\text{bkg})})$.

Let $i$ be the location of the bin corresponding to the correct depth of the scene point. This photon is only detected at $i$ if, in addition, no photon from the laser is detected at any prior bin. Since the laser photons only show up at bin $i$, constrained by depth, the probability of the photon showing up at any other bin is zero. However, in this conventional scenario, photons from ambient light could show up at any prior bin to $i$, pausing detection at bin $i$. Therefore, the probability that the photon from the laser is detected at the correct depth is $p_\text{sig} = (1 - e^{- (\Phi_\text{sig} + \Phi_\text{bkg})}) \ e^{-\Sigma^{i-1}_1 \Phi_\text{bkg}}$. 

The situation is different for ambient photons, which can arrive at any time instant  before photons from the $i^\text{th}$ bin arrive. We can write the probability that an ambient photon is detected at location $q$ as $p^q_\text{bkg} = (1 - e^{- \Phi_\text{bkg}}) \ e^{- \Sigma^{q-1}_1 \Phi_\text{bkg}}$. We can therefore write the SBR proportionality for the conventional imaging case as:

%\begin{equation}
%   SNR \propto \frac{(1 - e^{- \Phi_{sig}}) \ e^-{\Sigma^{i-1}_1} \Phi_{bkg}}{(1 - e^{- \Phi_{bkg}}) \ e^-{\Sigma^{i-1}_1} \Phi_{bkg}}.
%\end{equation}
\begin{equation}
   \text{SBR} \propto \frac{p_\text{sig}}{p_\text{bkg}} \propto \frac{(1 - e^{- (\Phi_\text{sig} + \Phi_\text{bkg})})) \ e^{-\Sigma^{i-1}_1 \Phi_\text{bkg}}}{\Sigma_{q=1}^{i} p^q_\text{bkg}}.
\end{equation}

% \input{Figs/FoveationResults_figure}
%%%%%%%%%%%%%%%%%%%%%%%%%%%%%%
% FileNames
\newcommand{\Fa}{00034}
\newcommand{\Fb}{00315}
\newcommand{\Fc}{00355}
\newcommand{\Fd}{00431}
\newcommand{\Fe}{01170}
% FileSubnames
\newcommand{\SubDS}{_DS}
\newcommand{\SubDSD}{_DSDepths}
\newcommand{\SubDSE}{_DSDepthsErrors}
\newcommand{\SubDSGS}{_DSDepthsGS}
\newcommand{\SubGTD}{nyu_depths}
\newcommand{\SubGTRGB}{nyu_rgb}
\newcommand{\SubFS}{_fullsim}
\newcommand{\SubFSE}{_fullsimErrors}
\newcommand{\SubFSGS}{_histGS}
\newcommand{\SubMD}{_memDepths}
\newcommand{\SubME}{_memErorrs}
\newcommand{\SubMGS}{_memGS}
\newcommand{\SubMono}{monocular_depth}

%Overleaf Paths, A = 60 B = 125 C = 250
\newcommand{\OPA}{Figs/S1_experiments/win60}
\newcommand{\OPB}{Figs/S1_experiments/win125}
\newcommand{\OPC}{Figs/S1_experiments/win250}
\newcommand{\OPGTRGB}{Figs/S1_experiments/RGB/}
\newcommand{\OPGTD}{Figs/S1_experiments/GT_D/}
\newcommand{\OPMono}{Figs/S1_experiments/Mono_D/}

\newcommand*{\incl}[1]{%
    {
    \includegraphics[width=0.143\linewidth,valign=m]{#1}
    }
}

\begin{figure*}[t!]

\includegraphics[width=\linewidth]{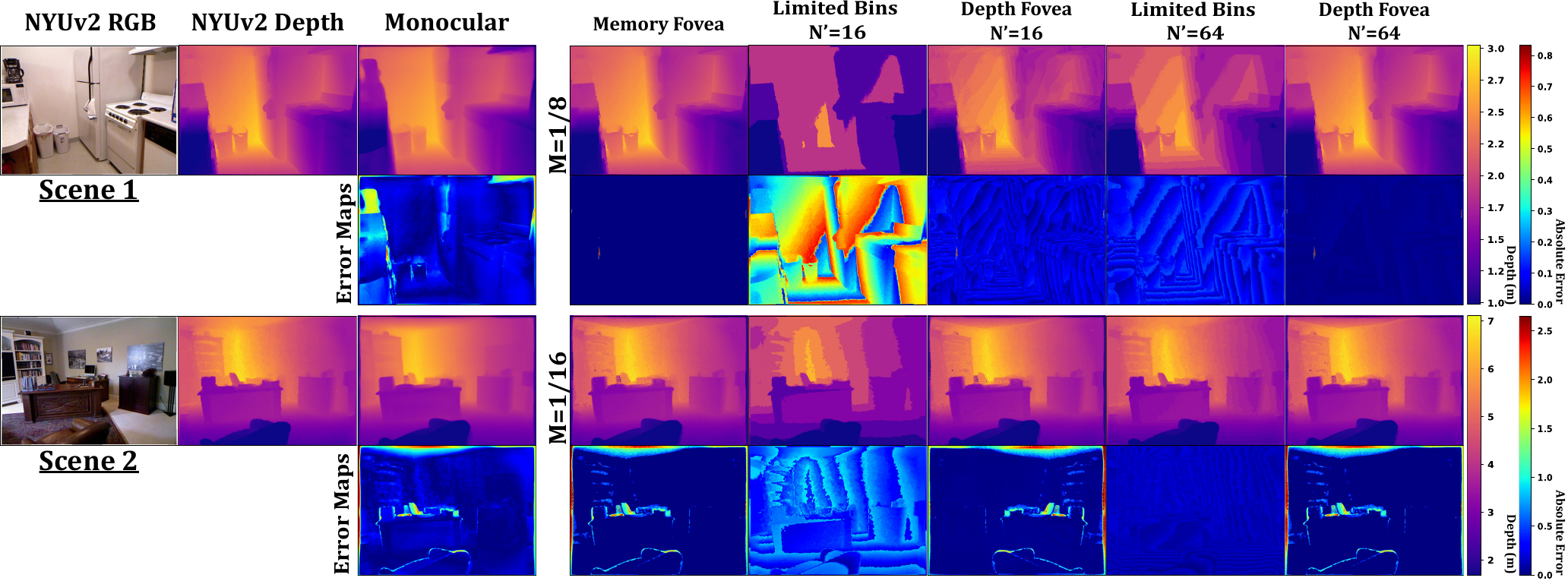}

 \caption{\textbf{Qualitative Comparison on NYUv2} Our memory and depth foveation techniques produce quality depth reconstructions with a fraction of the memory usage. Each row consists of the NYUv2 ground truth images, the monocular depth output from ZoeDepth, a simulated SPAD output with N$^\prime$ bins, and our foveation techniques. The rows show different combinations of M and N$^\prime$, where M is the number of bins in the foveated histograms, and N$^\prime$ is the limited number of bins used for depth foveation. Monocular estimation is just one method of obtaining a depth prior in a class of methods, in sec. \ref{sec:6opflow} and sec. \ref{sec:hardware} we show two more methods.} 
 % there are 3 algs, we show one
 \label{fig:compare_wSize_DS}
\end{figure*}

%Atul: This ratio looks like an SBR and not SNR. Are we only considering bin i for this equation? Because at bins j>i, the probability of detecting a bkg photon is also affected by the (earlier arriving) signal photon.

\smallskip
\noindent
\textbf{FoveaSPAD with Ambient Light:} We now consider both memory foveation and depth foveation where the foveated bins $N$ are given to us. In both these scenarios, we model the arrival of photons from both ambient and laser sources. 

\smallskip
\noindent
\textbf{Memory foveation:} Consider the foveated bins $N$, which we assume contain the bin with the histogram peak. Suppose the closest index for these bins is $j$. Then, the SBR increases, since the histogram sensitivity is unaffected by photons that impact the sensor before bin $j$. 

%\begin{equation}
%   SNR \propto \frac{(1 - e^{- \Phi_{sig}}) \ e^-{\Sigma^{i-1}_j} \Phi_{bkg}}{(1 - e^{- \Phi_{bkg}}) \ e^-{\Sigma^{i-1}_j} \Phi_{bkg}}.
%\end{equation}

\begin{equation}
   \text{SBR} \propto \frac{(1 - e^{- (\Phi_\text{sig} + \Phi_\text{bkg})}) \ e^{-\Sigma^{i-1}_j \Phi_\text{bkg}}}{\Sigma_{q=j}^{i} p^q_\text{bkg}}.
\end{equation}

In the extreme case, where we have perfect foveation, and $i=j$, then the terms for ambient light before bin $i$ become 1, 

\begin{equation}
   \text{SBR} \propto (1 - e^{- (\Phi_\text{sig} + \Phi_\text{bkg})}).
\end{equation}

%Atul: Again, not sure if this is SNR. Looks like SBR. This is also reminiscent of the async acquisition idea where we shift a subwindow to increase the probability of laser photon detection.

\noindent i.e. in other words, the effect of foveation is to remove the dependence on prior photon arrival for detection, since these no longer delay the measurement of photons at the $i$th bin. This ``perfect foveation" SBR term is dependent on the ratio of the strength of the laser and ambient signal directly and is not constrained by the binomial nature of SPAD photon capture. 

\smallskip
\noindent\textbf{Depth foveation:} Since we concentrate all $N$ bins into the foveation window, we are again susceptible to the binomial nature of SPAD photon capture. In addition, the bins are smaller to fit within the window, and as described in the non-ambient light section, the bin width is reduced as $\frac{M}{N}$. 

% Let the index of the first bin with the foveated window be $i_f$. 

We can write the probability that an ambient photon is detected at location $q$ as $p^q_\text{bkg} = (1 - e^{- \frac{M}{N}\Phi_\text{bkg}}) \ e^{- \Sigma^{q-1}_{1} \frac{M}{N}\Phi_\text{bkg}}$. The SBR proportionality also shows the effect of reduced signal strength as:

%\begin{equation}
%   SNR \propto \frac{(1 - e^{- \Phi_{sig}}) \ e^-{\Sigma^{i-1}_1} \Phi_{bkg}}{(1 - e^{- \Phi_{bkg}}) \ e^-{\Sigma^{i-1}_1} \Phi_{bkg}}.
%\end{equation}
\begin{equation}
   \text{SBR} \propto \frac{p_\text{sig}}{p_\text{bkg}} \propto \frac{(1 - e^{- (\frac{M}{N}(\Phi_\text{sig} + \Phi_\text{bkg}))}) \ e^{-\Sigma^{i-1}_{1} \frac{M}{N}\Phi_\text{bkg}}}{\Sigma_{q=1}^{i} p^q_\text{bkg}}.
\end{equation}

In summary, memory foveation increases SBR. While depth foveation has the same SBR as conventional capture, it improves depth resolution. It is this theory that motivates the remaining simulation results in the paper, where we explore different ways of creating depth and memory foveation for SPAD sensors.

%\begin{equation}
%   SNR \propto \frac{(1 - e^{- \frac{M}{N}\Phi_{sig}}) \ e^-{\Sigma^{i_f-1}_1} \frac{M}{N}\Phi_{bkg}}{(1 - e^{- \frac{M}{N}\Phi_{bkg}}) \ e^-{\Sigma^{i_f-1}_1} \frac{M}{N}\Phi_{bkg}}.
%\end{equation}

\section{SPAD Foveation from Monocular Depths \label{sec:4monofovea} }

With the imaging model defined, we proceed to our first experiment, demonstrating how our memory and depth foveation techniques can effectively work with a monocular depth prior.

Monocular depth estimation is inherently brittle due to biases in training datasets, whereas SPADs provide high-accuracy sensor measurements. In this section, we leverage the less accurate monocular depth to reduce the number of SPAD bins needed for capturing data, thereby saving memory and improving depth resolution.

\noindent\textbf{Simulation Details: } We conducted our simulations using the SPAD simulation framework provided in Gutierrez-Barragan et al. \cite{gutierrez2021itof2dtof,gutierrez2022compressive}, utilizing the code available on GitHub. While the simulations are initialized with RGBD datasets, all ``ground truth'' depth images presented in this paper result from SPAD simulation on full high-resolution histograms.

Monocular depth estimation algorithms use visual cues from 2D images to infer depth information and are trained on annotated datasets such as NYU Depth v2 \cite{SilbermanECCV12NYUV2} and KITTI \cite{Geiger2012CVPRKITTI}. We employed ZoeDepth \cite{bhat23mono}, a monocular depth estimator chosen for its performance and ability to produce metric depth estimates. The monocular depth is used to guide a foveation window consisting of $M$ bins in the histogram. The window size is a hyper-parameter, with larger sizes offering better accuracy at the cost of reduced efficiency.
 
For effective use of the monocular estimate as a prior, it must provide metric depth, and to enhance foveation performance, it needs to be scaled to match the scene. ZoeDepth fulfills the metric depth requirement, and we ensure compatibility with the dataset through appropriate scaling and bounding.

We chose a polynomial fit for scaling, observing that a majority of points in a randomly selected subset of the monocular output for the NYUv2 dataset exhibited a linear relationship. This scaling can be performed either locally, fitting the data to a specific scene, or generally across the dataset. In both cases, a small set of pixels is sampled at full histogram resolution, and the relationship between the monocular estimate and the SPAD estimate at these pixels is modeled. The fit is then applied to the entire monocular estimate, with bounds enforced for the minimum and maximum values across the dataset, which are 0m and 10m for NYUv2.

%We propose two methods to do this. One is a local method, which scales the monocular to a particular scene. In this method, at most 25 pixels in the scene are sampled at full bin resolution. The  We use a third order polynomial for this calibration, which changes depending on each scene. In contrast, in the second approach, which we call the general method, the SPAD pixels used for calibration are obtained at random from a training set of 10 scenes and this same polynomial is applied for every scene following this.  

We now describe our results shown in Fig. \ref{fig:compare_wSize_DS} and evaluated in Table \ref{table:fovea_eval} which are calibrated locally. The first two columns in the figure show the ground truth from the NYUv2 dataset. The depth is not simply the depth from the NYUv2 dataset, but the output of full-resolution SPAD simulation followed by the detection of the histogram peak. The third column shows the \textbf{scaled} monocular output. 

\noindent\textbf{Memory Foveation: } The fifth column in Fig. \ref{fig:compare_wSize_DS} shows our memory foveation results. Here, most bins are not used, saving memory for the same SNR. The foveated window is given at the right of the figure as a fraction of the original number of bins $N$, with $N$ set to 1000 bins for all experiments. The results are visually indistinguishable from ground truth, in some cases with a $\frac{1}{16}$ save in memory. In Table \ref{table:fovea_eval} we show the change in accuracy with these memory savings. Unsurprisingly, there is an inverse relationship between memory usage and depth error. 

\noindent\textbf{Depth Foveation: } In Fig. \ref{fig:compare_wSize_DS} the foveated window around the estimated monocular depth is packed with a limited number of bins. 
With no foveation, as in the fourth column, a limited number of bins $N^{'}$ are distributed over the entire SPAD volume. The depth foveation in the last column shows what happens when these limited number of bins are packed into the foveated window. Note that the depth resolution has increased from the limited bins case because the samples are placed within a foveated window where we expect to find the histogram peak. 
In Table \ref{table:fovea_eval}, entries with the same memory usage demonstrate the effects of depth foveation, where higher depth resolution consistently produces better results. These depth foveation outcomes are directly dependent on the memory foveation results, as both algorithms place fovea windows based on the same depth prior, with the depth foveation experiments having a lower depth resolution. Meaning, the memory foveation results establish a lower bound for the depth foveation error. Additionally, the limited bins case, which is not confined to a foveated window and thus reliant on a depth prior, shows that the error continues to decrease as depth resolution increases.

\definecolor{lightgray}{gray}{0.95}
\definecolor{midgray}{gray}{0.85}
\definecolor{gray}{gray}{0.75}

\begin{table*}
\caption{\textbf{Memory and Depth Foveation Evaluation - Local Scale} This table shows a quantitative comparison of RMSE and depth inlier metrics for different depth and memory foveation strategies for the NYUv2 dataset and a monocular estimation prior. For each memory foveation fraction, we vary the number of histogram bins in the foveated sub-window to achieve depth foveation. Metrics used from left to right: Root-mean-squared errror, Absolute $log_{10}$ error, Absolute Relative Error, $\delta < 1.25$, $\delta < 1.25^{2}$, $\delta < 1.25^{3}$}
\label{table:fovea_eval}
\begin{center}
%\centering
% \resizebox{1\columnwidth}{!}{
\tabcolsep=0.11cm
\begin{tabular}{|c|ccc|ccc||c|cccc|ccc|}
\hline
\textbf{\small{M}}  & \textbf{\small{RMSE}}$\downarrow$  &\textbf{log}$\bm{_{10}}$$\downarrow$ &\textbf{REL}$\downarrow$ & $\bm{\delta_1}$$\uparrow$& $\bm{\delta_2}$$\uparrow$ & $\bm{\delta_3}$$\uparrow$ & \textbf{N$^\prime$}  & \textbf{\small{RMSE}}$\downarrow$ & \textbf{\small{Lim. Bins}}$\downarrow$ &\textbf{log}$\bm{_{10}}$$\downarrow$ &\textbf{REL}$\downarrow$& $\bm{\delta_1}$$\uparrow$ & $\bm{\delta_2}$$\uparrow$ & $\bm{\delta_3}$$\uparrow$ \\
             (Fraction) & (m) & (m) && (\%)  & (\%) & (\%) &(Num. Bins)  & (m) & \textbf{\small{RMSE}} (m) &(m) &   & (\%)  & (\%) & (\%)   \\
\hline 
1/16  & 0.211 & 0.0106  & 0.0211 & 97.07  & 99.13 & 99.55 &
16  & 0.235  & 0.504 & 0.0173 &  0.0360 & 96.55 & 98.96 & 99.48 \\
 &  &  &  &  &  &  & 
32   & 0.211 & 0.250 & 0.0119 &  0.0241 & 97.1 & 99.14 & 99.55 \\
 &  &  &  &  &  &  & 
64 & 0.211 & 0.121 & 0.012 &  0.0242 & 96.44 & 99.01 & 99.54 \\
\hline
1/8 & 0.151 & 0.005 & 0.0109 & 98.36 & 99.42 & 99.79 & 
16 & 0.201 & 0.509 & 0.018 &  0.0418 & 97.87 & 99.26 & 99.71 \\
 &  &  &  &  &  & & 
32 & 0.184 & 0.250 & 0.011 & 0.0254 & 98.1 & 99.38 & 99.77 \\
 &  &  &  &  &  & &  
64  & 0.152 &  0.121 & 0.0064 &  0.0141 & 98.36 & 99.45 & 99.81 \\
\hline
1/4 & 0.117 & 0.0032 & 0.00686 & 99.24 & 99.57 & 99.79&  
16  & 0.221& 0.501  & 0.0326 &  0.0714 & 98.77 & 99.6 & 99.82 \\
  & &  &  &  &  & &  
32  & 0.166 & 0.2497 & 0.015 &  0.0355 & 99.15 & 99.59 & 99.82 \\
  &  &  &  & &  & &  
64  & 0.145 & 0.123 & 0.0087 &  0.0195 & 99.01 & 99.52 & 99.78 \\
\hline
% \bottomrule
\end{tabular}

\end{center}
\end{table*}
% \input{Figs/local_erros_mem_backup}

% Now describe columns 4 and6 in the figure. Say how the number of bins, when forced to go in uniform order produce quantization effects. These disappear when you place then in column 6 in the right places. Numerical results: refere to the table and say how you did on the numerical results.

\section{Spatio-Temporal SPAD Foveation}
\label{sec:5ST}

% FileNames
% \newcommand{\Fa}{00034}
% \newcommand{\Fb}{00315}
% \newcommand{\Fc}{00355}
% \newcommand{\Fd}{00431}
% \newcommand{\Fe}{01170}
\newcommand{\Ff}{00538}
\newcommand{\Fg}{01431}
\newcommand{\Fh}{01171}
\newcommand{\Fi}{00520}
% FileSubnames
% \newcommand{\SubDS}{_DS}
% \newcommand{\SubDSD}{_DSDepths}
% \newcommand{\SubDSE}{_DSDepthsErrors}
% \newcommand{\SubDSGS}{_DSDepthsGS}
% \newcommand{\SubGTD}{nyu_depths}
% \newcommand{\SubGTRGB}{nyu_rgb}
% \newcommand{\SubFS}{_fullsim}
% \newcommand{\SubFSE}{_fullsimErrors}
% \newcommand{\SubFSGS}{_histGS}
% \newcommand{\SubMD}{_memDepths}
% \newcommand{\SubME}{_memErorrs}
% \newcommand{\SubMGS}{_memGS}
% \newcommand{\SubMono}{monocular_depth}
\newcommand{\SubSp}{_sparse}
\newcommand{\SubQu}{_quant}

%Overleaf Paths, A = 60 B = 125 C = 250
% \newcommand{\OPA}{Figs/S1_experiments/win60}
% \newcommand{\OPB}{Figs/S1_experiments/win125}
% \newcommand{\OPC}{Figs/S1_experiments/win250}
% \newcommand{\OPGTRGB}{Figs/S1_experiments/RGB/}
% \newcommand{\OPGTD}{Figs/S1_experiments/GT_D/}
% \newcommand{\OPMono}{Figs/S1_experiments/Mono_D/}
\newcommand{\OPSTa}{Figs/ST_experiments/win60}
\newcommand{\OPSTb}{Figs/ST_experiments/win125/}
\newcommand{\OPSTc}{Figs/ST_experiments/win250/}

\newcommand*{\inclST}[1]{%
    {
    \includegraphics[width=0.152\linewidth,valign=m]{#1}
    }
}

%%%%%%%%%%%%%%%%%%%%%%%%%%%%%%

\begin{figure*}[!ht]
\centering
%%%%%%%%%%%%%%%%%%%%%%%%%%%%%%%%%%%%%%%%%%%%%%%%%%%%%%%%%%%%%%%%%%%%%%%%%%%%%%%%%%
% ROW 1
%%%%%%%%%%%%%%%%%%%%%%%%%%%%%%%%%%%%%%%%%%%%%%%%%%%%%%%%%%%%%%%%%%%%%%%%%%%%%%%%%%
\includegraphics[width=\linewidth]{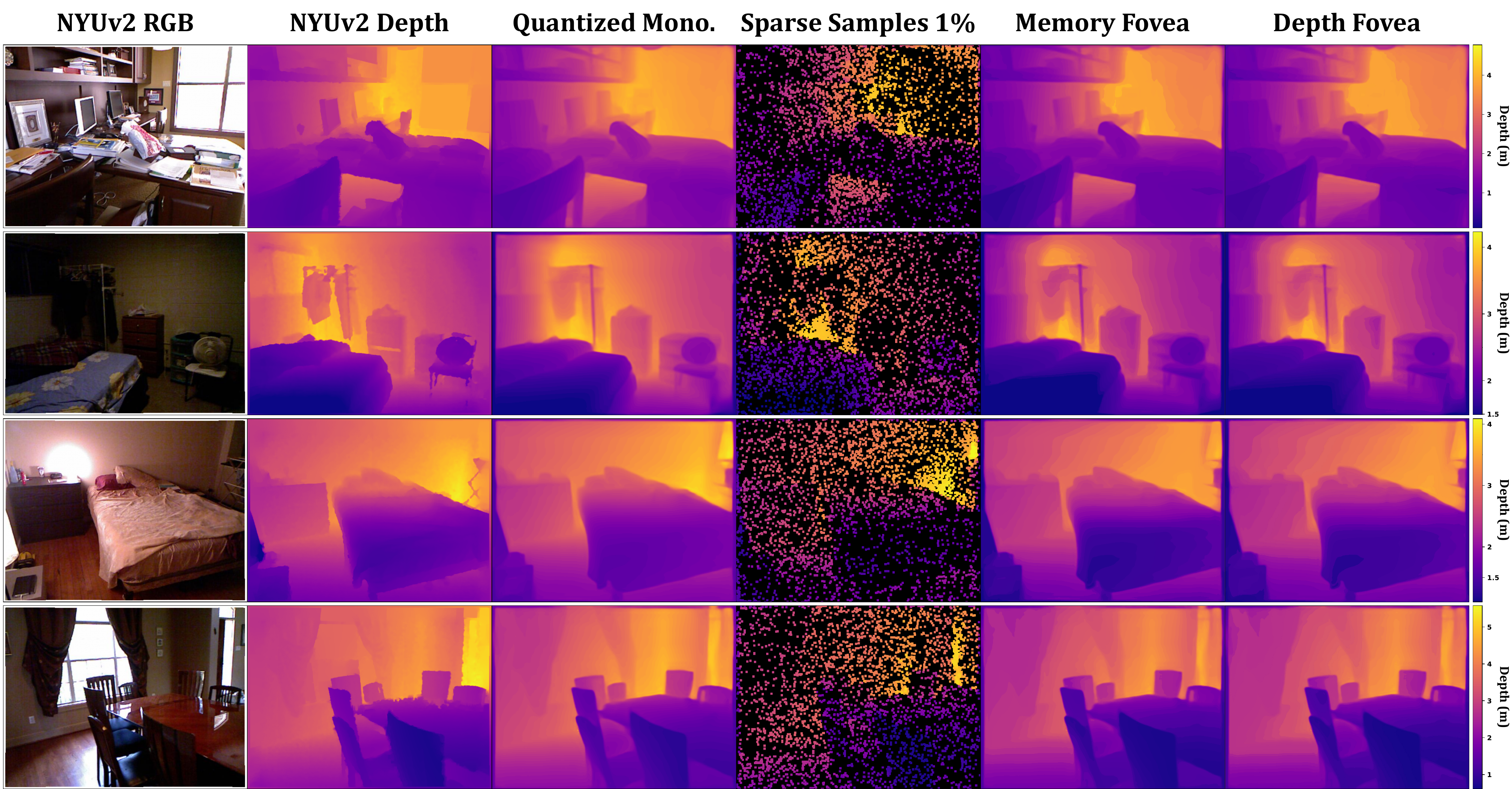}
 \caption{\textbf{Spatio-temporal foveation} The first two columns display the scene's color and ground truth depth. Using the quantized monocular depth in the third column, we select certain pixels in the fourth column. Processing only histograms at these locations with foveated windows generates results in the last column, indicating a 1548-fold reduction in memory usage. This is calculated by measuring memory allocation for full-res and spatio-temporal histograms. The results shown are with \textbf{M=1/16N} and \textbf{N$^\prime$ = 16}}

 %The first two columns show the color and ground truth depth of the scene. We use the quantized monocular depth in the third columns to select locations shown in the fourth column where the selected pixels are enlarged for visualization. Only histograms at these locations are processed, with foveated windows, to generate the results in the last column. These results show a 1548-fold decrease in memory usage, calculated by measuring memory allocation for full-res and spatio-temporal histograms.}

 %This factor was calculated by measuring the memory usage of the full resolution SPAD histogram and the spatio-temporal histogram, with the full resolution histogram as $(\text{ImageWidth}\times \text{ImageHeight},N)$ and the spatio-temporal histogram as $(\text{NumBuckets}\times \text{NumPixelsPerBucket},M)$}
\label{fig:ST}
\end{figure*}
\begin{table*}
\caption{\textbf{Spatio-Temporal Foveation Evaluation - Local Scale} Here we look at a quantitative comparison between the size of the foveation window (memory usage), the number of bins in depth foveation, and the number of total samples per the spatio-temporal algorithm.}
\label{table:spatiotemp_fovea_eval}
\begin{center}
%\centering
% \resizebox{1\columnwidth}{!}{
\tabcolsep=0.11cm
\begin{tabular}{|c||c|ccc|ccc||c|cccc|ccc|}
% \toprule
\hline

% TABLE 1 = QUANT16
\multirow{6}{*}{\rotatebox[origin=c]{90}{\textbf{Sparsity = 0.26 \% }}} &\textbf{\small{M}}  & \textbf{\small{RMSE}}$\downarrow$  &\textbf{log}$\bm{_{10}}$$\downarrow$ &\textbf{REL}$\downarrow$ & $\bm{\delta_1}$$\uparrow$& $\bm{\delta_2}$$\uparrow$ & $\bm{\delta_3}$$\uparrow$ & \textbf{N$^\prime$}  & \textbf{\small{RMSE}}$\downarrow$ & \textbf{\small{Lim. Bins}}$\downarrow$ &\textbf{log}$\bm{_{10}}$$\downarrow$ &\textbf{REL}$\downarrow$& $\bm{\delta_1}$$\uparrow$ & $\bm{\delta_2}$$\uparrow$ & $\bm{\delta_3}$$\uparrow$ \\
& (Fraction) & (m) & (m) & & (\%)  & (\%) & (\%) &(Num. Bins)  & (m) & \textbf{\small{RMSE}} (m) &(m) &   & (\%)  & (\%) & (\%)    \\

% & (Fraction)  & (m) & (m) & (\%) & (\%)  & (\%) & (\%) &(Num. Bins)  & (m) & \textbf{\small{RMSE}} (m) &(m) & (\%) & (\%)  & (\%) & (\%)   \\

% \hline 
\cline{2-16}
%mem fovea

&  1/16  & 0.39 & 0.06 & 0.124 & 84.901 & 97.054 & 99.429  &
%depth fovea
16  & 0.649 & 0.509 & 0.102 & 0.15 & 83.788 & 95.189 & 96.514 \\
& &  &  &  &  &  &  &
32   & 0.687 & 0.251  & 0.103 & 0.151 & 81.556 & 95.272 & 96.972 \\
&  1/8  & 0.392 & 0.068 & 0.137 & 80.154 & 94.812 & 99.046   &
%depth fovea
16  & 0.738 & 0.502 & 0.129 & 0.19 & 71.23 & 91.362 & 95.817 \\
& &  &  &  &  &  &  &
32   & 1.055 & 0.269 & 0.17 & 0.202 & 69.595 & 89.694 & 92.852 \\

%mem fovea
& 1/4 & 0.355 & 0.054 & 0.10 & 88.244 & 98.114 & 99.186 & 
%depth fovea
16 & 0.756 & 0.497 & 0.131 & 0.199 & 67.472 & 92.431 & 96.184 \\
& &  &  &  &  &  & &    
32 & 0.837 & 0.25 & 0.137 & 0.202 & 68.609 & 86.771 & 93.232\\

% 1/4 & 0.117 & 0.0032 & 6.86 & 99.24 & 99.57 & 99.79&  
% 16  & 0.221& 0.501  & 0.0326 &  7.14 & 98.77 & 99.6 & 99.82 \\
%   & &  &  &  &  & &  
% 32  & 0.166 & 0.2497 & 0.015 &  3.55 & 99.15 & 99.59 & 99.82 \\
%   &  &  &  & &  & &  
% 64  & 0.145 & 0.123 & 0.0087 &  1.95 & 99.01 & 99.52 & 99.78 \\
\hline
\hline

%SECOND TABLE = 32
\multirow{6}{*}{\rotatebox[origin=c]{90}{\textbf{Sparsity = 0.52 \% }}} &\textbf{\small{M}}  & \textbf{\small{RMSE}}$\downarrow$  &\textbf{log}$\bm{_{10}}$$\downarrow$ &\textbf{REL}$\downarrow$ & $\bm{\delta_1}$$\uparrow$& $\bm{\delta_2}$$\uparrow$ & $\bm{\delta_3}$$\uparrow$ & \textbf{N$^\prime$}  & \textbf{\small{RMSE}}$\downarrow$ & \textbf{\small{Lim. Bins}}$\downarrow$ &\textbf{log}$\bm{_{10}}$$\downarrow$ &\textbf{REL}$\downarrow$& $\bm{\delta_1}$$\uparrow$ & $\bm{\delta_2}$$\uparrow$ & $\bm{\delta_3}$$\uparrow$ \\
& (Fraction) & (m) & (m) & & (\%)  & (\%) & (\%) &(Num. Bins)  & (m) & \textbf{\small{RMSE}} (m) &(m) &   & (\%)  & (\%) & (\%)    \\

% \hline 
\cline{2-16}
%mem fovea
&  1/16  & 0.414 & 0.07 & 0.12 & 87.672 & 97.008 & 98.139 &
16  & 0.582 & 0.505 & 0.092 & 0.134 & 86.543 & 96.111 & 97.068 \\
& &  &  &  &  &  &  &
32   & 0.484 & 0.25 & 0.07 & 0.119 & 87.664 & 96.492 & 98.158 \\

& 1/8  & 0.387 & 0.051 & 0.108 & 87.292 & 99.162 & 99.919  &
%depth fovea
16  & 0.518 & 0.519 & 0.071 & 0.136 & 84.049 & 98.177 & 99.255 \\
& &  &  &  &  &  &  &
32 & 0.587 & 0.248 & 0.074 & 0.142 & 78.714 & 94.945 & 97.969 \\
%mem fovea
& 1/4 & 0.38 & 0.049 & 0.0996 & 90.254 & 96.965 & 98.365 & 
%depth fovea
16 & 0.734 & 0.518 & 0.121 & 0.184 & 74.702 & 93.679 & 96.225  \\
& &  &  &  &  &  & &
32 & 0.553 & 0.256 & 0.068 & 0.127 & 85.968 & 96.942 & 98.09\\

\hline
\hline
%THIRD TABLE TABLE = 64
\multirow{6}{*}{\rotatebox[origin=c]{90}{\textbf{Sparsity = 1.04 \% }}} &\textbf{\small{M}}  & \textbf{\small{RMSE}}$\downarrow$  &\textbf{log}$\bm{_{10}}$$\downarrow$ &\textbf{REL}$\downarrow$ & $\bm{\delta_1}$$\uparrow$& $\bm{\delta_2}$$\uparrow$ & $\bm{\delta_3}$$\uparrow$ & \textbf{N$^\prime$}  & \textbf{\small{RMSE}}$\downarrow$ & \textbf{\small{Lim. Bins}}$\downarrow$ &\textbf{log}$\bm{_{10}}$$\downarrow$ &\textbf{REL}$\downarrow$& $\bm{\delta_1}$$\uparrow$ & $\bm{\delta_2}$$\uparrow$ & $\bm{\delta_3}$$\uparrow$ \\
& (Fraction) & (m) & (m) & & (\%)  & (\%) & (\%) &(Num. Bins)  & (m) & \textbf{\small{RMSE}} (m) &(m) &   & (\%)  & (\%) & (\%)    \\

% \hline 
\cline{2-16}
%mem fovea
&  1/16  & 0.288 & 0.039 & 0.0855 & 94.214 & 99.582 & 99.935 &
%depth fovea 16?? 16?? 16?? 16?? 16?? ?? ?? ?? ?? ?? ?? ?? ?? ??
16  & 0.364 & 0.508 & 0.048 & 0.0959 & 93.693 & 99.248 & 99.646 \\
& &  &  &  &  &  &  &
32   & 0.412 & 0.254 & 0.051 & 0.0933 & 93.048 & 98.179 & 99.145 \\

& 1/8  & 0.313 & 0.04 & 0.0881 & 91.782 & 99.443 & 99.841  &
%depth fovea
16  & 0.386 & 0.495 & 0.056 & 0.111 & 90.719 & 99.057 & 99.474\\
& &  &  &  &  &  &  &
32   & 0.432 & 0.257 & 0.053 & 0.106 & 89.662 & 98.472 & 99.276 \\

%mem fovea
& 1/4 & 0.274 & 0.035 & 0.0786 & 94.264 & 99.045 & 99.875 & 
%depth fovea
16 & 0.471 & 0.503 & 0.072 & 0.148 & 82.311 & 97.104 & 98.821  \\
& &  &  &  &  &  & &    
32 & 0.399 & 0.25 & 0.063 & 0.111 & 91.482 & 97.966 & 98.643\\

% 1/4 & 0.117 & 0.0032 & 6.86 & 99.24 & 99.57 & 99.79&  
% 16  & 0.221& 0.501  & 0.0326 &  7.14 & 98.77 & 99.6 & 99.82 \\
%   & &  &  &  &  & &  
% 32  & 0.166 & 0.2497 & 0.015 &  3.55 & 99.15 & 99.59 & 99.82 \\
%   &  &  &  & &  & &  
% 64  & 0.145 & 0.123 & 0.0087 &  1.95 & 99.01 & 99.52 & 99.78 \\
\hline
% \bottomrule
\end{tabular}

\end{center}
\end{table*}

The previous section seeks to reduce the SPAD histogram bottleneck by reducing the number of bins to examine per-pixel with a monocular estimate prior. This section aims to improve these savings by incorporating spatial foveation. By exploiting depth coherencies and applying foveated windows to a small selection of pixels we show an order of magnitude increased bandwidth savings.

%However, we can also save processing spatially by exploiting depth coherencies and applying foveated windows to a small selection of pixels. 

Foveated LiDAR systems \cite{tasneem2018dirrectionally,pittaluga2020towards,adaptivelidarstanford} can place samples onto depth edges and recover the rest of the scene, post-capture, through algorithmic estimation such as deep guided upsampling or gradient-based reconstruction. Similarly, here, we place samples \emph{across} depth edges and, rather than use an algorithm, we use the SPAD measurement to provide correct depths in redundant areas.

\noindent\textbf{Quantized Sampling: } Our approach to spatial sampling begins by quantizing the prior through thresholding, resulting in digitized regions that we refer to as `buckets.' We make the \emph{assumption} that the values within each quantized bucket are redundant. From each bucket, we randomly select pixels and use the SPAD to measure these points in the scene, applying memory foveation in the process. These measurements provide a sparse depth map, which we subsequently sort and quantize based on the buckets defined by the depth prior.

%To do this, we threshold the depth prior, quantizing the prior into discrete buckets, with the \emph{assumption} that the values in a quantized bucket are redundant. This allows us to reduce the amount of SPAD measurements to just a few within the bucket.

%Taking the depth from the SPAD simulation we then fill the bucket region it came from, and repeat this for the rest of the buckets. 

In Fig. \ref{fig:ST} we show examples of our approach, where the first two columns show the scene and ground truth depths. The third column is a quantized version of the monocular depth estimation, where the number of quantized buckets is 64. For each of these buckets, we picked 50 points at random and recovered the SPAD depths of these points. Note that these transients were also foveated in time, using the method described in the previous section. 
The fourth column in Fig. \ref{fig:ST} depicts exactly those points in the SPAD camera that were sampled, with the number of bins sampled at $\frac{1}{16}$ of the original histogram. This is a factor of $1548$ memory savings, compared to the ground truth measurement, with depth results in the last column. These efficiencies are evaluated in Table \ref{table:spatiotemp_fovea_eval}. 

\begin{comment}
- recall that we had the calibrated monocular driving foveation. we can quantize this into buckets (we use the word buckets to differentiate from bins). For each bucket, we only need to process one scene point, and we use the depth of that scene point for the whole scene.

- describe your figure here, how many scenes you did, describe the different columsn in your figure. write down the numerical results that you got for this data. 
\end{comment}

\section{Optical Flow Driven SPAD Foveation}
\label{sec:6opflow}
In previous sections, we focused on static scenes. However, one of the key advantages of using SPAD arrays is their fast capture speed, making them ideal for dynamic environments, such as when mounted on a vehicle. In this section, we demonstrate how our techniques can be applied to moving scenes by utilizing optical flow to guide the foveation process.

Consider a SPAD sensor on a moving platform, say an autonomous vehicle, where high-frame rate and efficient depth capture are important \cite{lee2023caspi,beer2018spad}. 
The foveation algorithm described in the previous section analyses pixels in each frame, reducing the bins in the histogram that need to be processed.
Here we consider an approach to reduce the computation even further, using temporal information by transferring foveation information from previous frames to subsequent frames.

Consider a sequence of frames containing both depth and reflectance information from a scene. Assume that the depth in the first frame is reconstructed at high quality, such as from full-resolution SPAD histograms. Now, for a subsequent frame, we can calculate optical flow between the frames (color or grayscale), producing a vector $(u,v)$ for each pixel at a given time $t$. These vectors satisfy the brightness consistency principle, meaning that $I(x + u\cdot\delta t, y + v\cdot \delta t, t + \delta t) = I(x,y,t)$.  holds true. We use the depth information from the previous frame to guide the positioning of the foveating window in the current frame, by warping the previous frame based on the vector $(u,v)$. Although the object may move and the histogram peak will shift from frame to frame, it will remain within a nearby range, allowing a window of pixels to recover the histogram peak in the current frame.

However, optical flow is never perfect, often having errors at the edges of a frame. Further, these propagate incorrect depths through time, since our optical flow method only considers the depths in the previous frame. To remove this error, we compare the distribution of the photons under a foveated region to that from a noise floor. If they match, we ignore the erroneous optical flow, and recompute depth from the full histogram. In practice, this is done by thresholding the values in the foveated window. 

In Fig. \ref{fig:OFD}, we show some optical flow results. Please see the supplementary video for all of our video results. These were created on the CARLA simulator \cite{dosovitskiy2017carla} and the results show two street scenes with ground truth depths. We found the native optical flow in CARLA to be noisy, and so we used OpenCV's in-built optical flow estimator. The third and fourth columns show first the incorrect results from optical flow, and our method to detect these regions, shown in red. The optical flow driven depth foveation results are shown in the last column. Calculating errors using a running average across all video frames reveals compounding errors over time. In the first scene, at $\frac{1}{10}N$, RMSE and SSIM are 101.9m and 0.530, and at $\frac{1}{4}N$, 38.6m and 0.884. In the second scene, RMSE and SSIM are 0.164m and 0.87 for both $\frac{1}{10}N$ and $\frac{1}{4}N$.

% \noindent \textbf{Time Delay:} In dynamic scenes, unlike static ones, time delay becomes a potential issue. The process of calculating optical flow may introduce a delay between capturing the prior frame and the current one, which could lead to increased errors. Additionally, the proposed error correction method in this section relies on resampling pixels, which could introduce time-delay artifacts if the frame rate is not sufficiently high. However, with a fast enough capture speed, as provided by SPAD arrays, these delays become negligible, maintaining the accuracy of the output depth map.
%Calculating errors with a running average over all the frames in a video, we see that the errors compound over time. For the first scene, we found at $\frac{1}{10}N$, RMSE and SSIM values of 101.9m and 0.530, and at $\frac{1}{4}N$ 38.6m and 0.884. While for the second scene, we observe RMSE and SSIM values of 0.164m and 0.87 for both  $\frac{1}{10}N$ and $\frac{1}{4}N$.
% OF NUMBERS %%%%%% Small Paragraph

% FileNames
\newcommand{\OFa}{/depth_}

% FileSubnames
% \newcommand{\SubDS}{_DS}
% \newcommand{\SubDSD}{_DSDepths}
% \newcommand{\SubDSE}{_DSDepthsErrors}
% \newcommand{\SubDSGS}{_DSDepthsGS}
% \newcommand{\SubGTD}{nyu_depths}
% \newcommand{\SubGTRGB}{nyu_rgb}
% \newcommand{\SubFS}{_fullsim}
% \newcommand{\SubFSE}{_fullsimErrors}
% \newcommand{\SubFSGS}{_histGS}
% \newcommand{\SubMD}{_memDepths}
% \newcommand{\SubME}{_memErorrs}
% \newcommand{\SubMGS}{_memGS}
% \newcommand{\SubMono}{monocular_depth}

%Overleaf Paths, A = 60 B = 125 C = 250
\newcommand{\OPARGB}{Figs/OFD_experiments/Scene0/RGB}
\newcommand{\OPAGTD}{Figs/OFD_experiments/Scene0/GT}
\newcommand{\OPAFTT}{Figs/OFD_experiments/Scene0/FoveaThreshold/010}
\newcommand{\OPAFTQ}{Figs/OFD_experiments/Scene0/FoveaThreshold/025}

\newcommand{\OPBRGB}{Figs/OFD_experiments/Scene7/RGB}
\newcommand{\OPBGTD}{Figs/OFD_experiments/Scene7/GT}
\newcommand{\OPBFTT}{Figs/OFD_experiments/Scene7/FoveaThreshold/010}
\newcommand{\OPBFTQ}{Figs/OFD_experiments/Scene7/FoveaThreshold/025}

\newcommand{\OPAOOT}{Figs/OFD_experiments/Scene0/OFonly/010}
\newcommand{\OPAOOQ}{Figs/OFD_experiments/Scene0/OFonly/025}

\newcommand{\OPBOOT}{Figs/OFD_experiments/Scene7/OFonly/010}
\newcommand{\OPBOOQ}{Figs/OFD_experiments/Scene7/OFonly/025}

\newcommand*{\inclOF}[1]{%
    {
    \includegraphics[width=0.175\linewidth,valign=m]{#1}
    }
}

%%%%%%%%%%%%%%%%%%%%%%%%%%%%%%

\begin{figure*}
\setlength{\tabcolsep}{0.01cm}

\centering
%%%%%%%%%%%%%%%%%%%%%%%%%%%%%%%%%%%%%%%%%%%%%%%%%%%%%%%%%%%%%%%%%%%%%%%%%%%%%%%%%%
% ROW 1
%%%%%%%%%%%%%%%%%%%%%%%%%%%%%%%%%%%%%%%%%%%%%%%%%%%%%%%%%%%%%%%%%%%%%%%%%%%%%%%%%%
    \begin{tabular}{c c c c c c}
        \textbf{Carla RGB} & \textbf{Carla GT Depth} & \textbf{Optical Flow Driven}  & \textbf{Optical Flow Error} & \textbf{Ours} \\
%%%%%%%%%%%%% 1ST SET %%%%%%%%%%%%        
        \inclOF{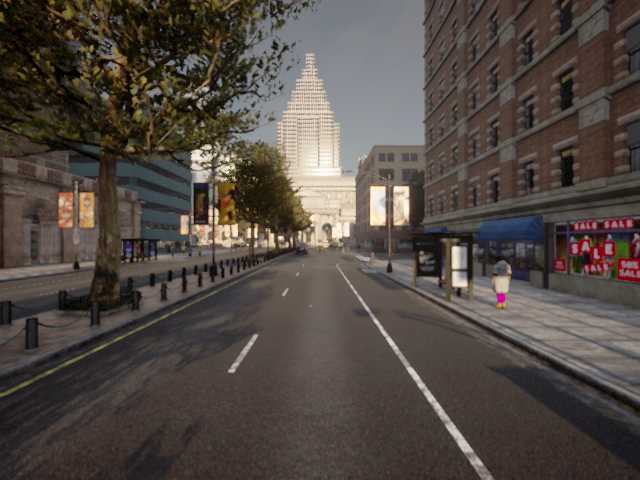} &
        \inclOF{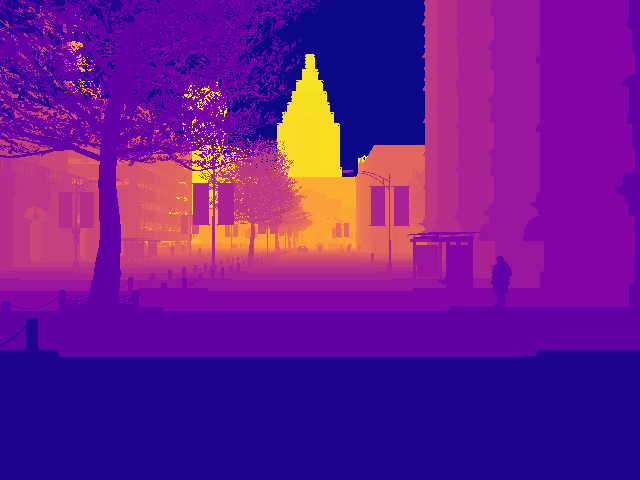} & 
        \inclOF{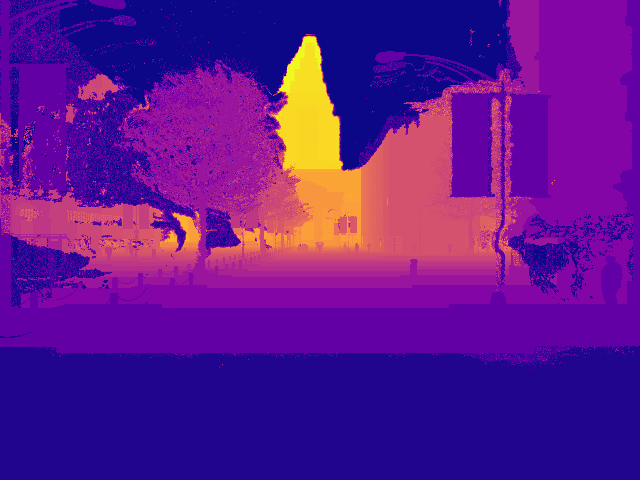} & 
        \inclOF{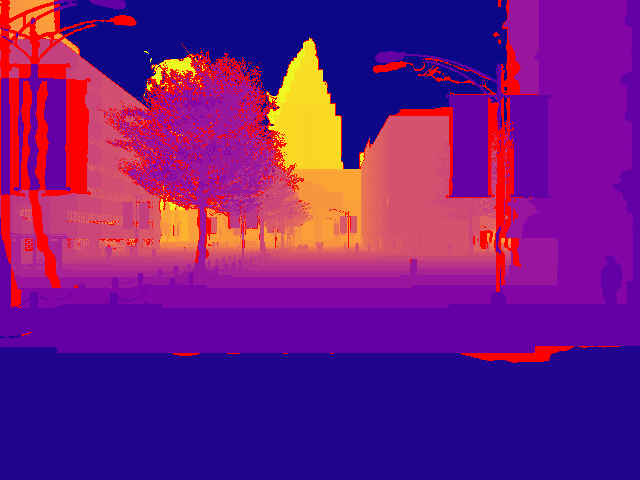} & 
        \inclOF{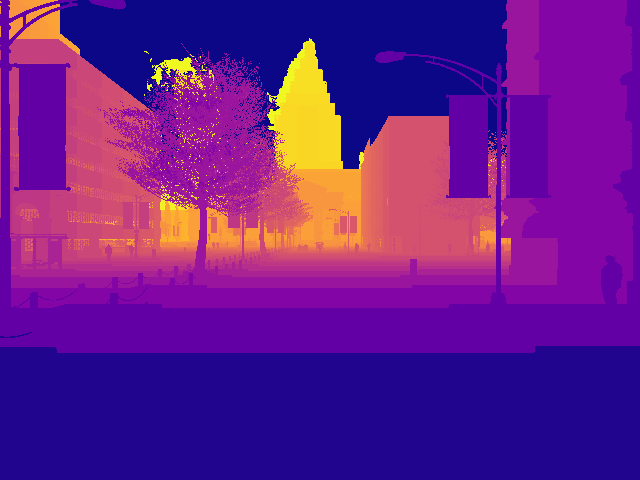} & 
        \textbf{\normalsize{M = 1/4N}}\\
        
        \inclOF{\OPARGB/rgb_0005.png} &
        \inclOF{\OPAGTD\OFa0005.png} & 
        \inclOF{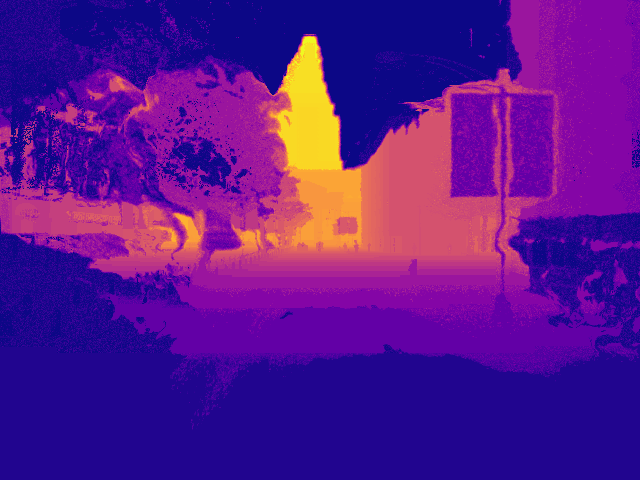} & 
        \inclOF{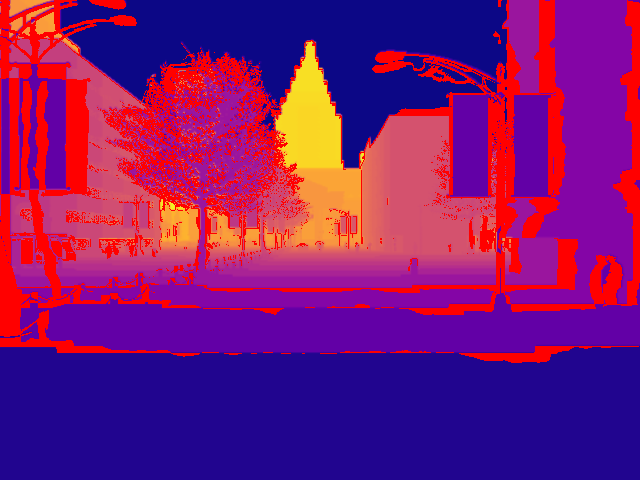} & 
        \inclOF{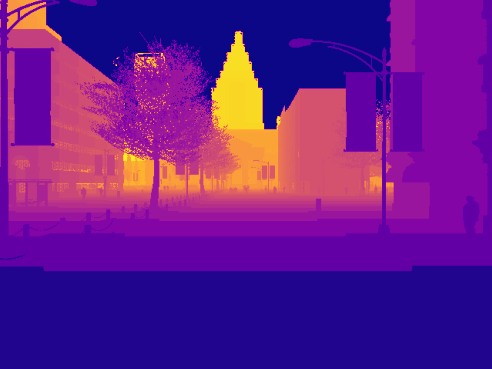} & 
        \textbf{\normalsize{M = 1/10N}}\\
        \\
        %%%%%%%%%%%%%% 2ND SET %%%%%%%%%%
        \inclOF{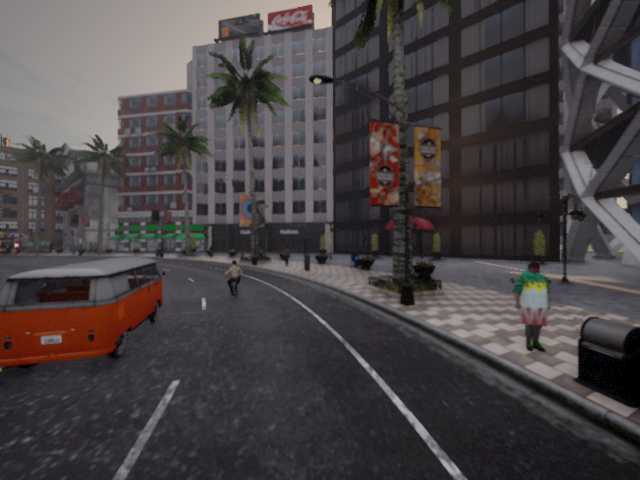} &
        \inclOF{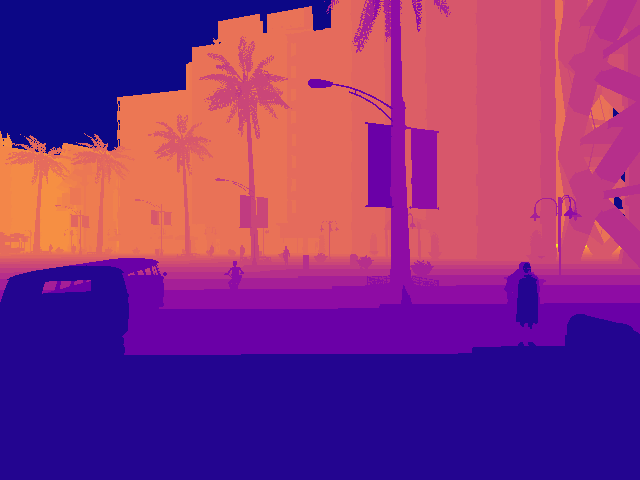} & 
        \inclOF{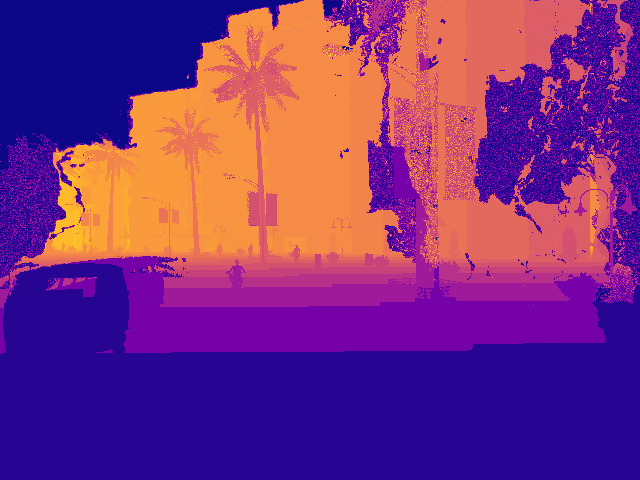} & 
        \inclOF{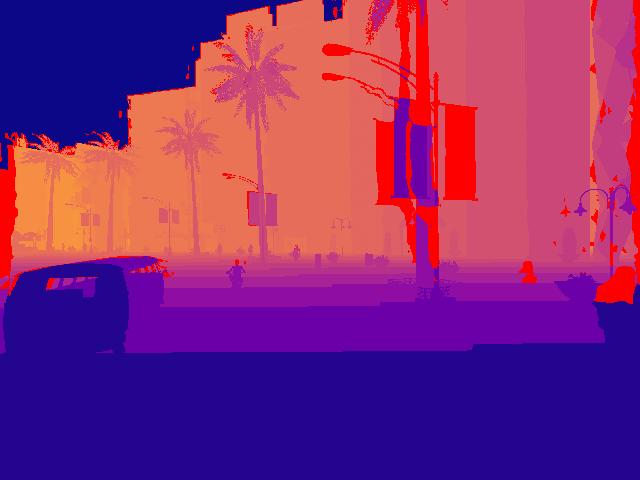} & 
        \inclOF{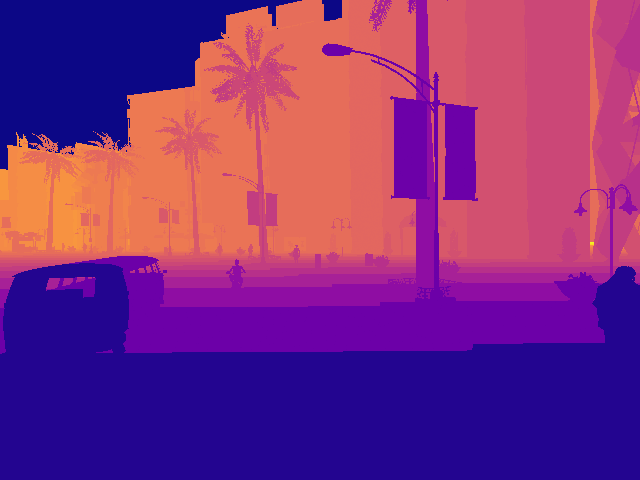} & 
        \textbf{\normalsize{M = 1/4N}}\\

        \inclOF{\OPBRGB/rgb_0097.png} &
        \inclOF{\OPBGTD\OFa0097.png} & 
        \inclOF{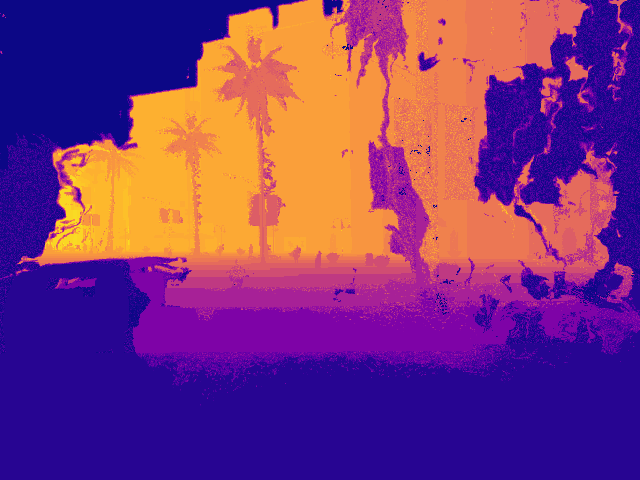} & 
        \inclOF{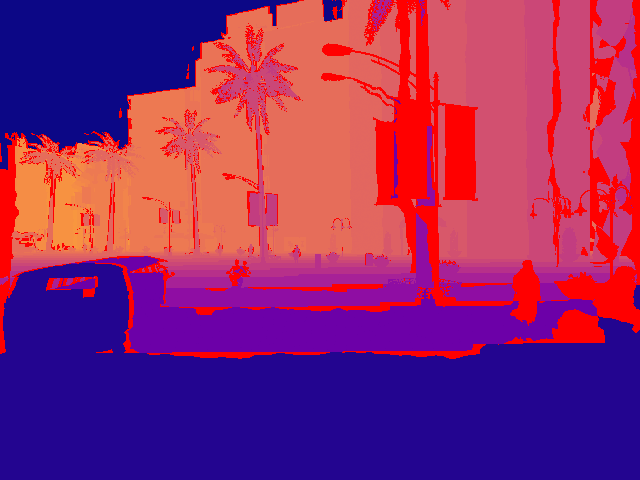} & 
        \inclOF{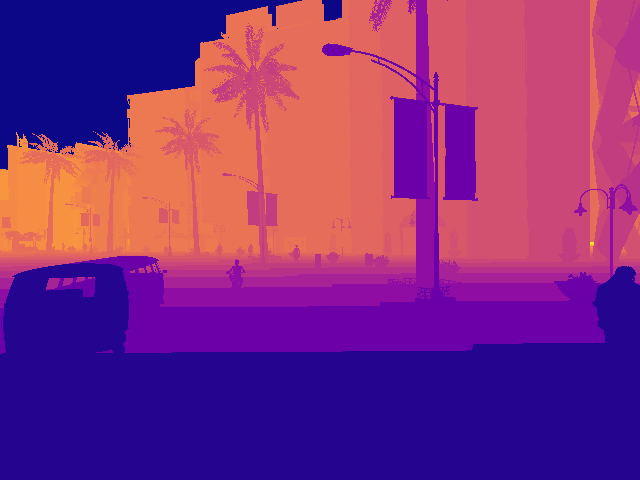} & 
        \textbf{\normalsize{M = 1/10N}}\\

    \end{tabular}

 \caption{\textbf{Optical Flow Driven Foveation} Here we see our optical flow driven SPAD foveation using the Carla simulator whose color and ground-truth depth are shown in the first two columns. Directly using optical flow, as shown in the third column, creates errors that propagate over time. We correct for the optical flow error by detecting those pixels whose foveated windows are close to the noise floor. The last column shows the final optical flow driven foveated depth at different window sizes. Please see the supplementary for video results. } %%Write MORE
\label{fig:OFD}
\end{figure*}

\section{Hardware Emulation Results}
\label{sec:hardware}

\begin{figure*}[!t]
 \centering
 \includegraphics[width=1.00\textwidth]{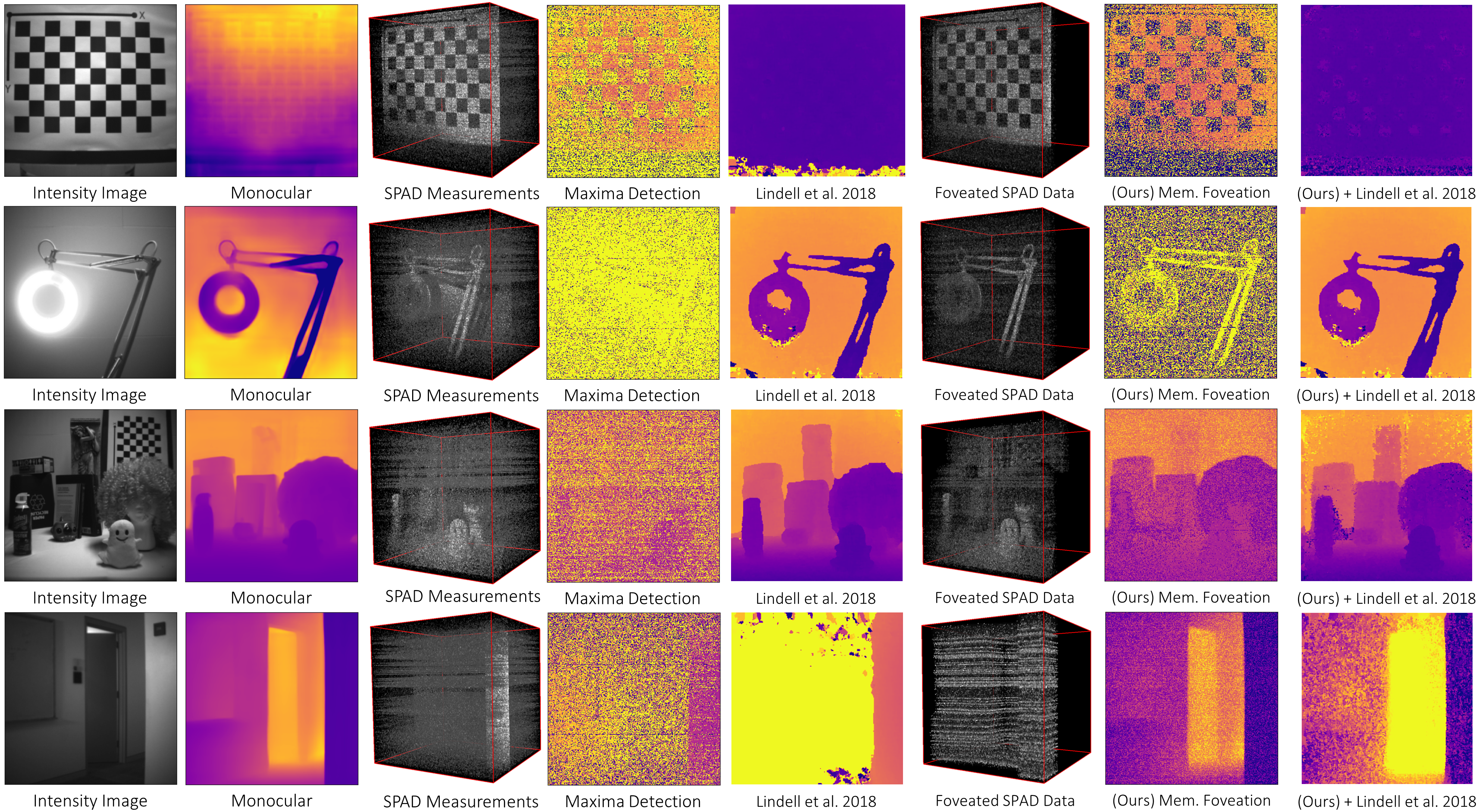}
 \caption{\textbf{Hardware emulation results for scenes from Lindell et al. \cite{lindell2018single}.}  (Column 1) The Lindell dataset consists of monochrome images captured by a camera co-aligned with the SPAD sensor that captures photon data cubes.
 (Column 2) We obtain monocular depth maps from these monochrome images.
 (Column 3) Raw photon data cube without foveation shows a ``cloud'' of background photon detections.
 (Column 4) Maxima detection on low SBR photon clouds leads to unusable depth maps.
 (Column 5) The CNN-based algorithm of Lindell et al. improves depth map reconstruction.
 (Column 6) Our approach relies on memory foveation in a 1/4th size sub-window around an estimate of the true depth obtained from monocular depth maps. Observe that the photon data cubes are less noisy.
 (Column 7) Even a simple max-estimator provides better depth map estimates after foveation.
 (Column 8) Providing foveated clouds to the CNN denoiser of Lindell et al. further improves reconstructions.
 \label{fig:lindell_fig}}
\end{figure*}

In this section, we present hardware emulation results for depth and memory foveation using SPAD data captured using real hardware. The goal of hardware emulation study is to de-risk future in-pixel implementations of foveation algorithms. We use datasets by Lindell et al. \cite{lindell2018single} and Gutierrez-Barragan et al.\cite{gutierrez2022compressive} from prior sources \cite{gupta2019asynchronous,gupta2019photon}.

%Bad Table No Cookie
%\input{Figs/opticalFlow_table}
\subsection{Using Monocular for Memory Foveation}
We'll start by showcasing how our memory foveation technique works on the dataset by Lindell et al. \cite{lindell2018single} by using monocular as a prior. The Lindell dataset consists of scenes under different ambient illumination conditions captured using a linear SPAD pixel array \cite{burri2017linospad} co-aligned with a monochrome camera that captures intensity images.

We use these intensity images to obtain a monocular depth prior. Because the performance of monocular estimation networks is dependent on the dataset, we perform a calibration step by using the ``elephant'' scene in the dataset to define a global scaling function.
We place foveation windows of 1/4th the total temporal extent of the full histograms centered around these scaled monocular depth estimates for each pixel.

Memory foveation improves the overall SBR, in a scene-adaptive manner, by focusing on regions of the spatio-temporal photon cube where signal photons arrive.
Comparing columns 3 and 6 in Fig. ~\ref{fig:lindell_fig}, foveated SPAD measurement cubes show fewer background photon detections, with clear 3D object structure in the photon cubes.
Depth estimates are improved even with a simple maxima-detection approach --- observe that the lamp is barely visible in the non-foveated maxima-detection-based depth map in column 5, but is visible after memory foveation in column 7.
Running memory foveated measurements through the denoising algorithm of Lindell et al. further improves the depth map, as seen in the last column of Fig.~\ref{fig:lindell_fig}.

\begin{figure*}[!t]
 \centering
 \includegraphics[width=0.9\textwidth]{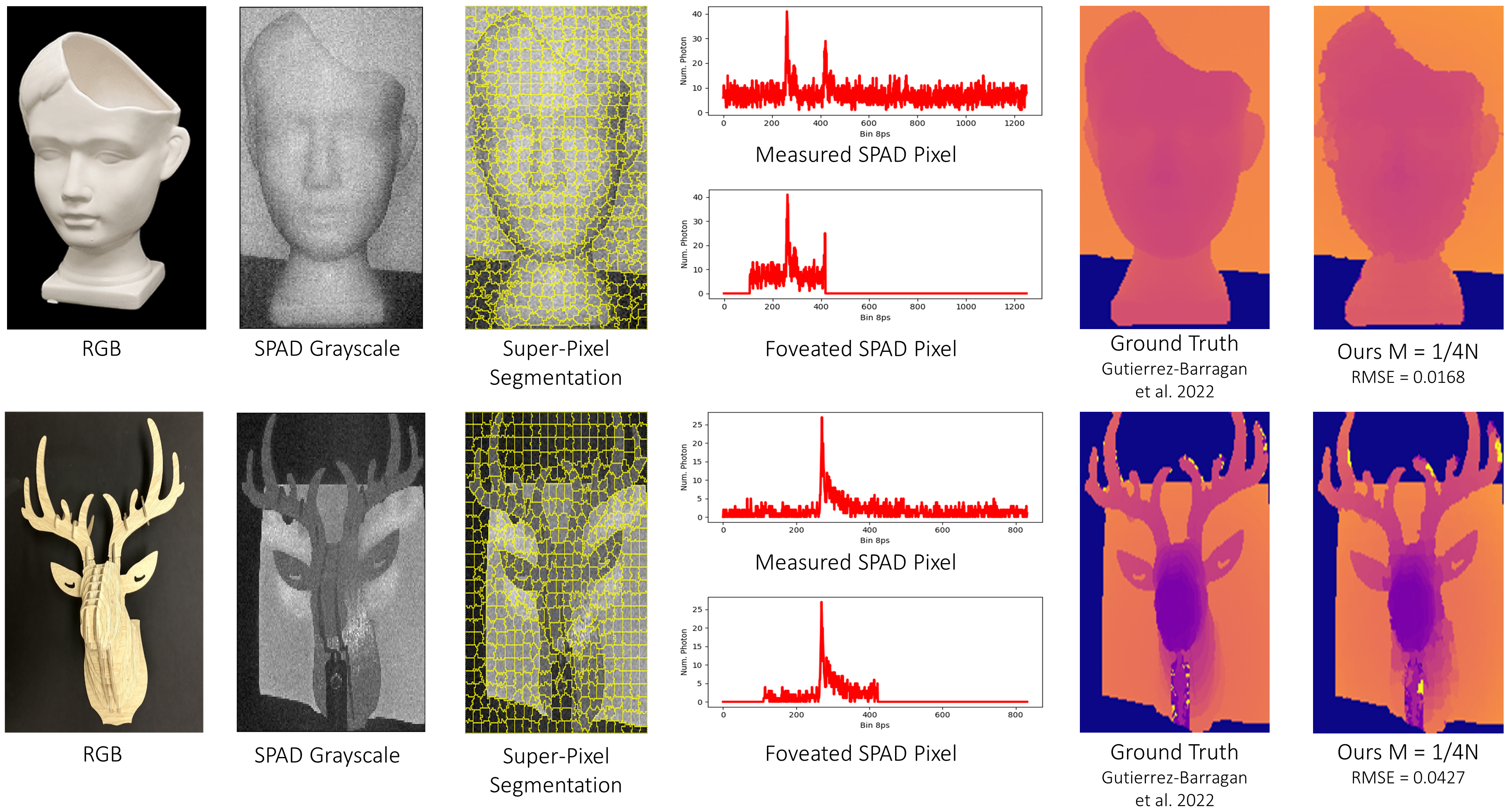}
 \vspace{-0.1in}
 \caption{\textbf{Hardware emulation results for scenes without co-aligned monochrome camera \cite{gutierrez2022compressive}.}  (Column 1) RGB images of the ``face-vase'' and ``reindeer'' scenes shown for visualization.
 (Column 2) A pseudo-intensity image is estimated by accumulating photon counts for each pixel.
 (Column 3) Pseudo intensity maps are converted into superpixel representations, and a single pixel in each superpixel is used for measuring complete histograms.
 (Column 4) The peak location of the chosen pixel is used to apply foveation windows of 1/4th the total temporal extent for the remaining pixels in each superpixel.
 (Column 5) Ground truth depth maps obtained using matched filtering.
 (Column 6) Our result requires 64$\times$ less memory per pixel for $>99\%$ of the pixels in these scenes.
 \label{fig:atul_fig} }
 \vspace{-0.1in}
\end{figure*}

\subsection{A Different Approach to Spatio-Temporal Foveation}
To illustrate the flexibility of our foveation techniques and their independence from external sensors as a prior, we propose an alternative spatio-temporal method, which we apply to two scenes from the Gutierrez-Barragan et al. dataset \cite{gutierrez2022compressive}, for which there is no co-located camera. The dataset is captured using a single-pixel point scanned SPAD detector co-aligned with a pulsed laser. Fig ~\ref{fig:atul_fig} shows the results of the alternate approach for the single object ``face-vase'' and ``reindeer'' scenes, with the RGB images shown in column 1 for visualization purposes.

\noindent\textbf{SuperPixels: }Because there is no intensity map captured in the dataset, we instead obtain a pseudo-intensity map by summing the raw photon data cubes along the temporal axis for each pixel. In a real hardware implementation, this process would be achieved by utilizing a counter in each SPAD pixel, a feature commonly available in existing commercial SPAD arrays. We then run a superpixel algorithm \cite{snic_cvpr17} on the pseudo-intensity maps  to obtain coarse segmentations of the scene, as shown in column 3. For each superpixel segment, we capture a complete (non-foveated) histogram of the centroid pixel. By identifying the true peak location in this histogram, we can then foveate within a 1/4th sub-window centered around this peak for all remaining pixels in the superpixel segment, reducing the overall bandwidth requirement per pixel by a factor of 64.

In the ``face-vase'' scene, with a spatial resolution of $174 \times 154$ pixels, the segmentation reduces the data to 473 superpixels. Similarly, the ``deer'' scene, originally at  $204\times 116$ pixels, is reduced to 515 superpixels.
This reduction translates to a 3/4 reduction in memory requirement for approximately 99.98\% pixels in both scenes. Examples of foveated histograms in column 4 show that the laser impulse response function has a non-ideal shape which departs significantly from the commonly assumed Gaussian shape used in simulation studies.
(The second peak is likely due to optical inter-reflections in the hardware setup).
Yet, our method is able to produce reliable depth maps (columns 5 and 6).

\begin{figure}
    \centering
    \includegraphics[width=0.95\linewidth]{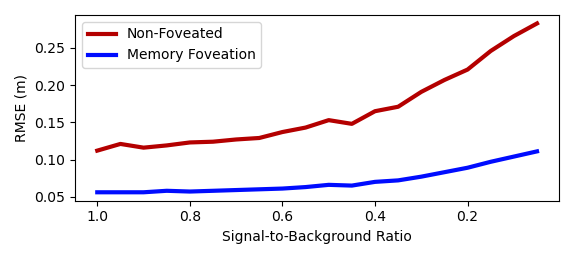}
    \caption{\textbf{Effect of increasing background illumination.} The conventional (non-foveated) depth map quality degrades more rapidly as background illumination increases. Using memory foveation allows reliable depth map recovery for the ``deer'' scene for a wider range of SBR levels.}
    \label{fig:sbr_error}
\end{figure}

We also examine the impact of reconstruction error under increasing background noise for the "deer" scene. As shown in Fig.~\ref{fig:sbr_error}, foveation allows for the accurate selection of the correct depth peak, even in the presence of strong background illumination, thereby expanding the operable SBR range in practice.

\begin{figure*}
    \centering
    \includegraphics[width=\textwidth]{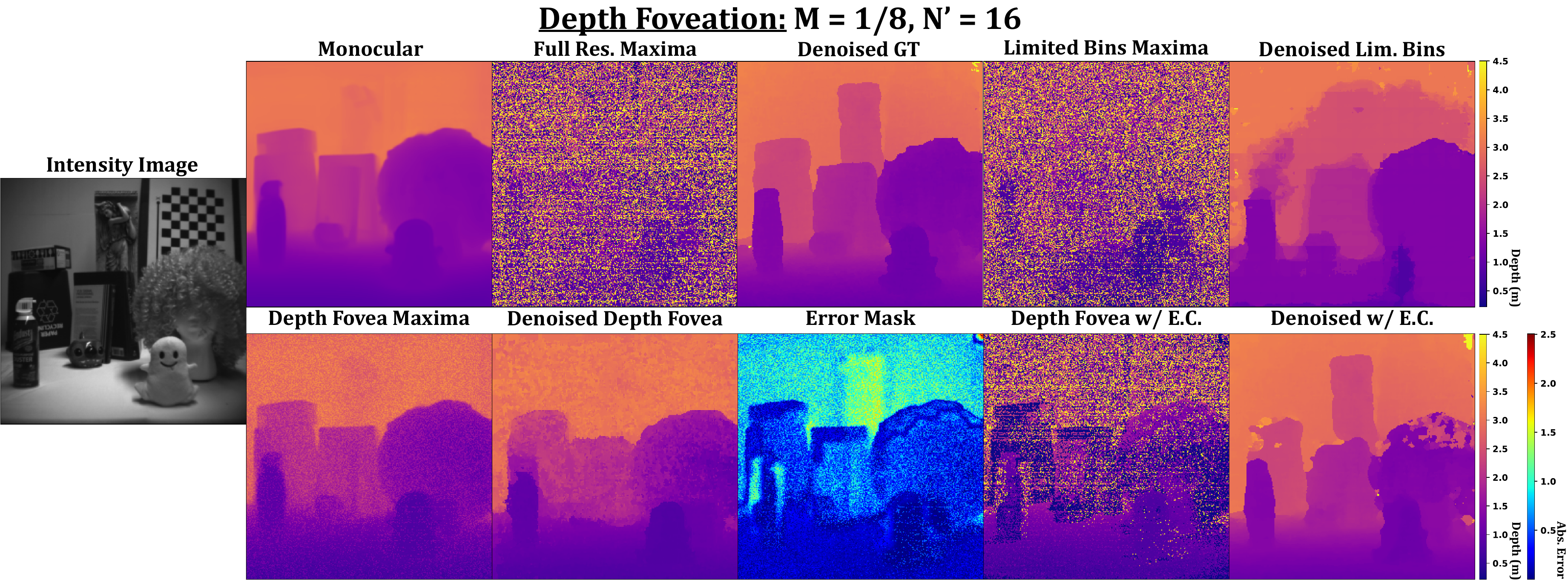}
    \caption{\textbf{Additional Results: Depth Fovea.} This figure demonstrates the application of the depth foveation technique described in Sec. \ref{sec:4monofovea} to the Lindel dataset, along with the error correction technique presented in the supplementary material. A window size of M = 1/8 and a bin count of N' = 16 were used. The results were subsequently processed using the sensor fusion denoising network \cite{lindell2018single}.}
    \label{fig:addResults_DepthFovea}
\end{figure*}

\begin{figure}
    \centering
    \includegraphics[width=0.4\textwidth]{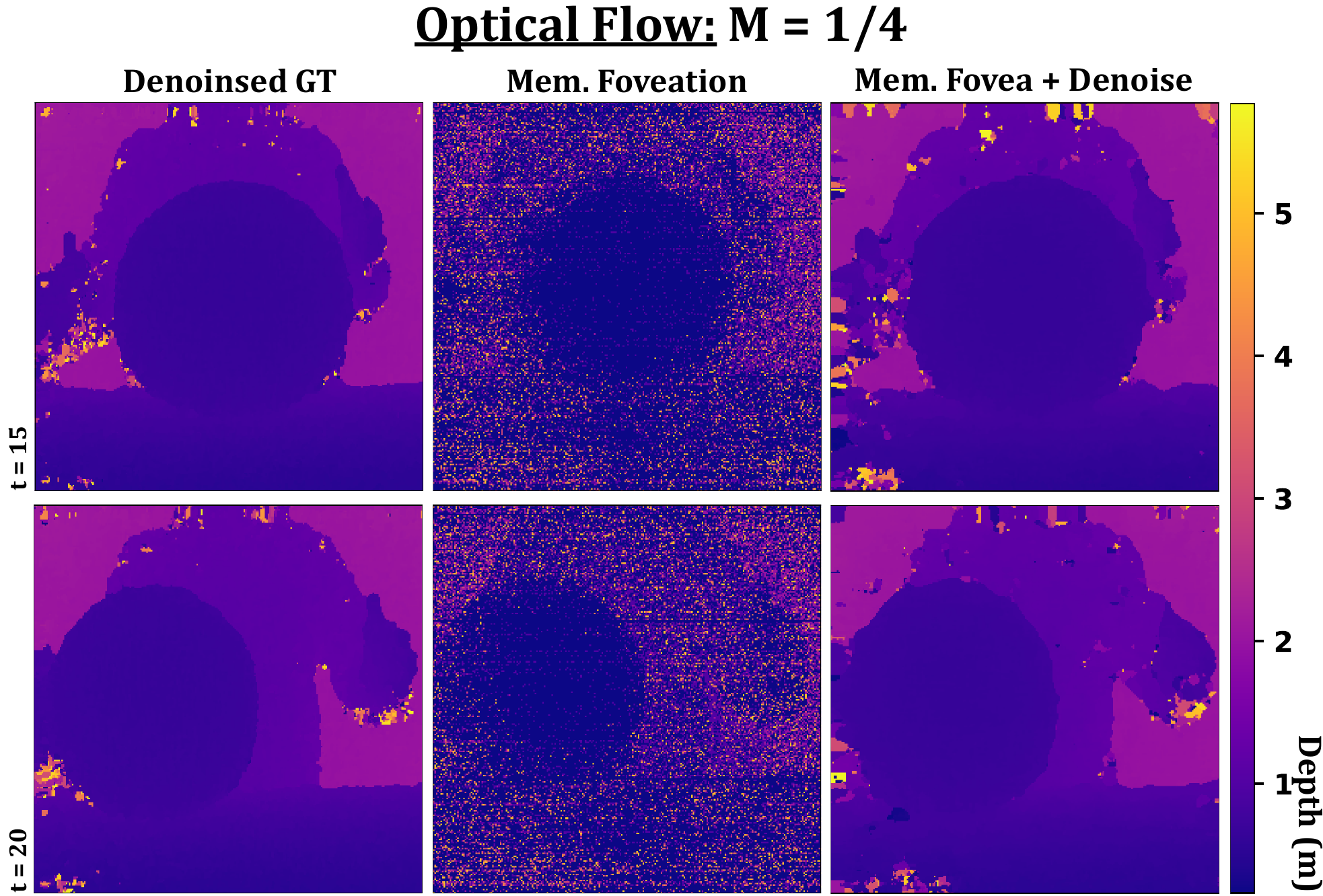}
    \includegraphics[width=0.4\textwidth]{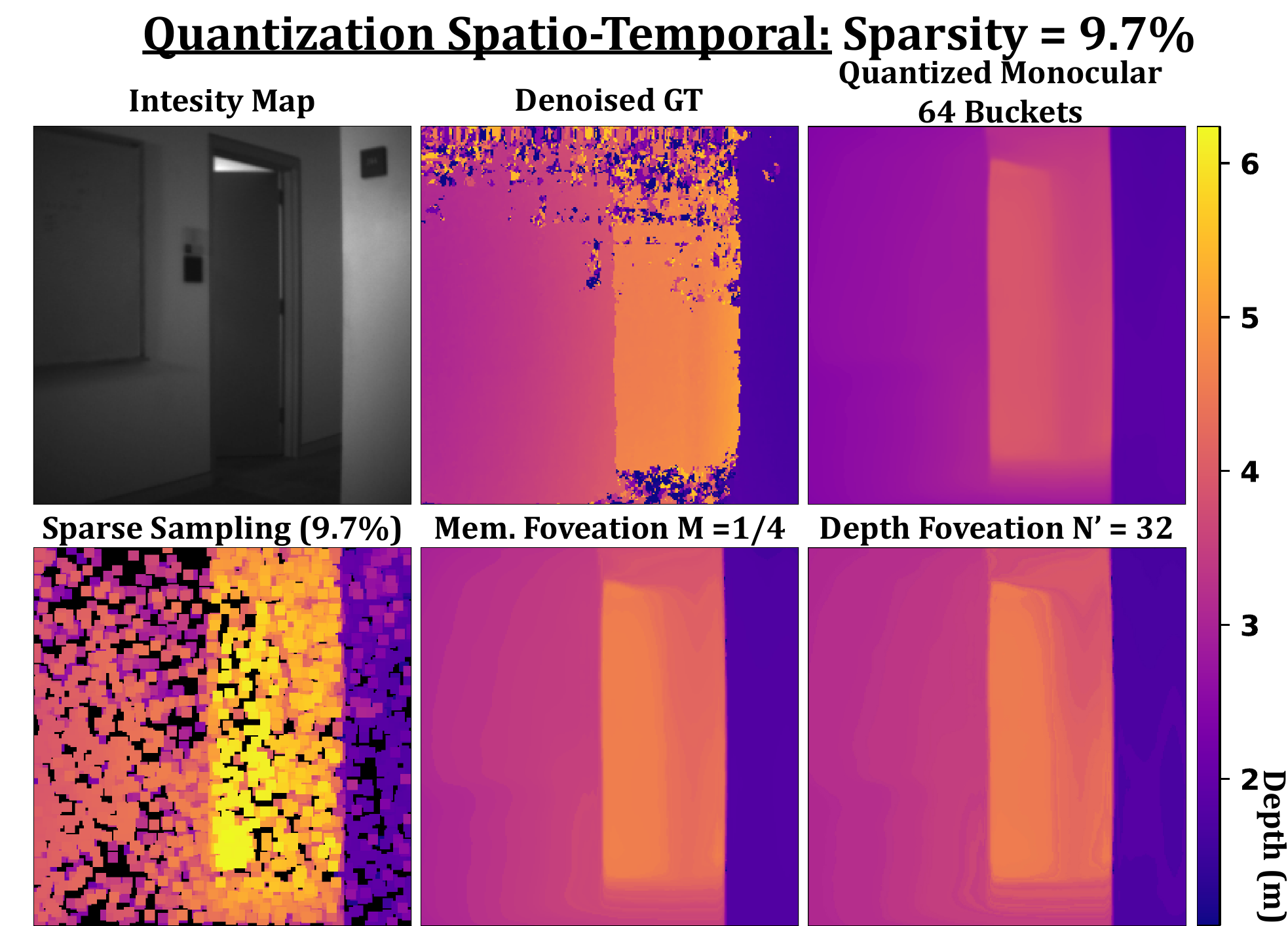}
    \caption{\textbf{Additional Results: Optical Flow and Quantization Spatio-Temporal.} This figure illustrates the application of the techniques described in Sec. \ref{sec:6opflow} and Sec. \ref{sec:5ST} to the Lindel dataset. The top portion showcases our optical flow algorithm on the "roll" scene. The first column displays the denoised ground truth, followed by the optical-flow-driven memory foveation result using maxima detection, and finally the denoised memory foveation result. The bottom portion of the figure presents our quantization spatio-temporal foveation technique, utilizing 9.7\% sampling to mitigate the high levels of noise and the abundance of pixels with no photon counts in the scene.}
    \label{fig:addResults_OFST}
\end{figure}

\section{Limitations and Discussion}
\label{sec:limit}
\noindent \textbf{Worst Case Stochastic Limits:}
We explored the limitations of our approach by analyzing the worst-case scenario where depth is incorrectly detected due to various errors, such as monocular depth calibration issues, ambient light interference, and global effects like multipath inter-reflections. We characterized these errors using a probabilistic framework. Specifically, we defined the probability \( p_{\text{gt}} \) as the chance that a detected photon originates from the laser i.e. single-bounce photons, \( p_{\text{multipath}} \) as the probability of multipath photon detection, and \( p_{\text{floor}} \) as the probability of spurious peaks due to sensor noise. The overall probability of accurate depth detection is given by

\begin{equation}
p_{\text{gt}} (1 - p_{\text{gt}} p_{\text{multipath}})^{M-1} p_{\text{floor}},
\end{equation}

where \( M \) is the number of foveated bins. We further derived the probability \( p_{\text{worst}} \) for the worst-case scenario, where none of the \( S \) pixels detect the correct depth, expressed as 

\begin{equation}
p_{\text{worst}} = (1 - p_{\text{gt}} (1 - p_{\text{gt}} p_{\text{multipath}})^{M-1} p_{\text{floor}})^S.
\end{equation}

Through optimization, we identified two conditions that lead to this worst-case scenario, linked to specific relationships between \( p_{\text{gt}} \), \( p_{\text{multipath}} \), and \( M \). 
\begin{itemize}
    \item The first condition occurs when \( p_{\text{gt}} = \frac{1}{p_{\text{multipath}}} \). This situation arises when the probability is 1 for every bin to contain both direct photons from the laser and photons that have undergone multipath effects, indicating a degenerate scene, such as one made entirely of mirror-like surfaces.
    
    \item The second condition occurs when \( p_{\text{gt}} = \frac{1}{M \cdot p_{\text{multipath}}} \). This scenario implies that the number of foveated bins \( M \) and the probability of multipath effects \( p_{\text{multipath}} \) must satisfy this relationship, under the constraint that \( 0 \leq p_{\text{gt}} \leq 1 \). This suggests that it is possible to avoid the worst-case scenario by adjusting the number of bins \( M \) for scenes with specific global illumination characteristics.
\end{itemize}

In order to illustrate the findings of this analysis, consider a toy example with a number of bins \( M = 1000 \) and pronounced multipath effects, such as \( p_{\text{multipath}} = 0.1 \). In the worst case, the probability of depth recovery would be significantly hindered \( p_{\text{gt}} = 0.01 \), but can be improved by changing the number of bins \(M\) at the cost of depth resolution. The detailed derivations of these results are provided in the supplementary material.

\noindent \textbf{Quality of depth priors:} Our algorithms can enable memory-efficient SPAD sensing while maintaining depth accuracy. However, our method strongly relies on the accuracy of the depth prior. If the prior is incorrect, our algorithms may produce errors, highlighting the importance of robust error correction mechanisms. We can correct for such errors by trading off efficiency. We show one example error mask in the supplementary which can be used to drive corrections, such as a larger foveation window (using the entire span of the transient in the extreme case).

\begin{figure}
    \centering
    \includegraphics[width=1.0\linewidth]{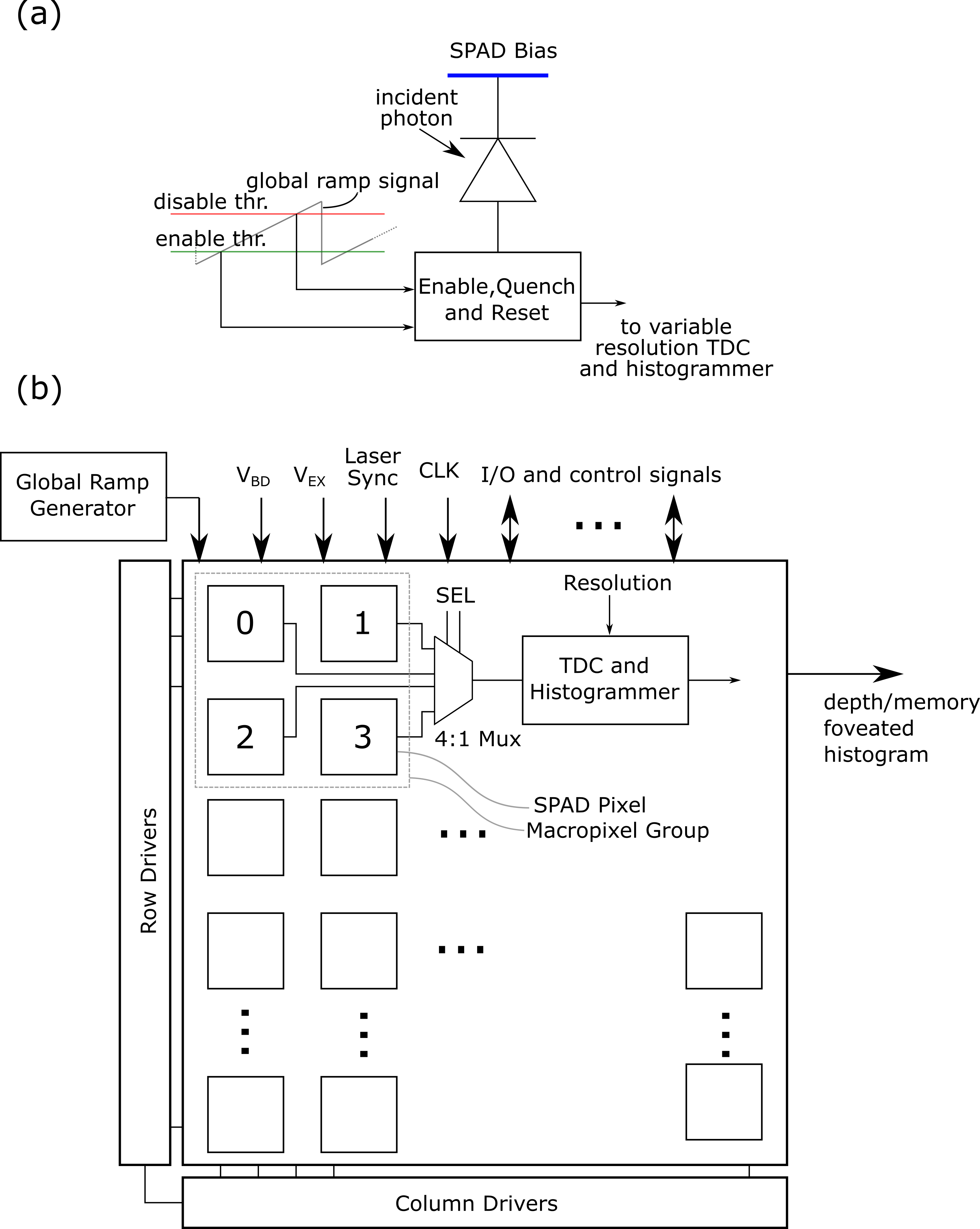}
    \caption{
    \textbf{Future pixel and array designs for foveated single-photon 3D imaging.} 
    (a) A speculative pixel design where individual SPADs are gated on or off based on thresholds set with respect to a linear ramp signal. 
    Pixels only need to store the thresholds; the ramp signal is generated externally.
    (b) A possible array of SPAD pixels with per-pixel gating.
    Observe that the ramp signal is generated globally, simplifying pixel design.
    Variable-resolution TDCs and histogrammers are shared by small pixel neighborhoods (e.g., $2 \times 2$ multiplexed ``macropixels") to improve fill factor.}
    \label{fig:pixeldesign}
\end{figure}

\smallskip
\noindent\textbf{Hardware complexity:}
A key limitation of our approach is the lack of available hardware that fully supports our algorithms, necessitating more complex pixel architectures and driving up costs. Each SPAD pixel in the 2D array requires a programmable gate, along with a variable TDC and histogrammer, which increases the complexity and expense of the hardware. This presents a significant challenge to the widespread adoption and practical implementation of our method. In Fig.~\ref{fig:pixeldesign}, we propose a potential array design with per-pixel gating capability, where a global ramp generator provides individualized on/off thresholds for each pixel. To enhance the fill factor, the TDC and histogrammer are shared among groups of neighboring pixels, forming ``macropixels''.

We believe the next generation of programmable and software-defined SPAD cameras \cite{ardelean2023computational,sundar2023sodacam} will be key enablers for in-pixel and on-chip implementation of memory- and energy-efficient foveated sensing schemes. 
As SPAD cameras become low-cost and widely available \cite{callenberg2021low}, the integration of in-pixel foveated sensing algorithm proposed here will reduce memory consumption while maintaining depth accuracy, or alternatively, provide more accurate depth estimates without increasing memory usage.

%% Add more limitations
% No available hardware
%  - More Complex pixel architecture
%  - Expensive
% Dependency on the prior
    % If the prior is wrong, then we can "miss".
    % Error Correction as an area of interest.

%\appendices
%\input{appendix}
% Any acknowledgments to only be included in camera ready

\section*{Acknowledgments}
Authors Folden and Koppal gratefully acknowledge the partial support of the National Science Foundation (1942444 and 2330416) and the Office of Naval Research (N000142312429 and N000142312363). Author Ingle was supported in part by National Science Foundation Award 2138471.

\bibliographystyle{IEEEtran}
\bibliography{references}

% Generated by IEEEtran.bst, version: 1.14 (2015/08/26)
\begin{thebibliography}{10}
\providecommand{\url}[1]{#1}
\csname url@samestyle\endcsname
\providecommand{\newblock}{\relax}
\providecommand{\bibinfo}[2]{#2}
\providecommand{\BIBentrySTDinterwordspacing}{\spaceskip=0pt\relax}
\providecommand{\BIBentryALTinterwordstretchfactor}{4}
\providecommand{\BIBentryALTinterwordspacing}{\spaceskip=\fontdimen2\font plus
\BIBentryALTinterwordstretchfactor\fontdimen3\font minus \fontdimen4\font\relax}
\providecommand{\BIBforeignlanguage}[2]{{%
\expandafter\ifx\csname l@#1\endcsname\relax
\typeout{** WARNING: IEEEtran.bst: No hyphenation pattern has been}%
\typeout{** loaded for the language `#1'. Using the pattern for}%
\typeout{** the default language instead.}%
\else
\language=\csname l@#1\endcsname
\fi
#2}}
\providecommand{\BIBdecl}{\relax}
\BIBdecl

\bibitem{vaswani2017attention}
A.~Vaswani, N.~Shazeer, N.~Parmar, J.~Uszkoreit, L.~Jones, A.~N. Gomez, {\L}.~Kaiser, and I.~Polosukhin, ``Attention is all you need,'' \emph{Advances in neural information processing systems}, vol.~30, 2017.

\bibitem{selvaraju2017grad}
R.~R. Selvaraju, M.~Cogswell, A.~Das, R.~Vedantam, D.~Parikh, and D.~Batra, ``Grad-cam: Visual explanations from deep networks via gradient-based localization,'' in \emph{Proceedings of the IEEE international conference on computer vision}, 2017, pp. 618--626.

\bibitem{Xiong1993defocus}
Y.~Xiong and S.~Shafer, ``Depth from focusing and defocusing,'' in \emph{Proceedings of (CVPR) Computer Vision and Pattern Recognition}, June 1993, pp. 68 -- 73.

\bibitem{lindell2018single}
D.~B. Lindell, M.~O'Toole, and G.~Wetzstein, ``Single-photon 3d imaging with deep sensor fusion,'' \emph{ACM Transactions on Graphics (ToG)}, vol.~37, no.~4, pp. 1--12, 2018.

\bibitem{gutierrez2022compressive}
F.~Gutierrez-Barragan, A.~Ingle, T.~Seets, M.~Gupta, and A.~Velten, ``Compressive single-photon 3d cameras,'' in \emph{Proceedings of the IEEE/CVF Conference on Computer Vision and Pattern Recognition}, 2022, pp. 17\,854--17\,864.

\bibitem{hutchings2019reconfigurable}
S.~W. Hutchings, N.~Johnston, I.~Gyongy, T.~Al~Abbas, N.~A. Dutton, M.~Tyler, S.~Chan, J.~Leach, and R.~K. Henderson, ``A reconfigurable 3-d-stacked spad imager with in-pixel histogramming for flash lidar or high-speed time-of-flight imaging,'' \emph{IEEE Journal of Solid-State Circuits}, vol.~54, no.~11, pp. 2947--2956, 2019.

\bibitem{zhang201830}
C.~Zhang, S.~Lindner, I.~M. Antolovi{\'c}, J.~M. Pavia, M.~Wolf, and E.~Charbon, ``{A 30-frames/s, $252 \times 144 $ SPAD Flash LiDAR With 1728 Dual-Clock 48.8-ps TDCs, and Pixel-Wise Integrated Histogramming},'' \emph{IEEE Journal of Solid-State Circuits}, vol.~54, no.~4, pp. 1137--1151, 2018.

\bibitem{heide2018sub}
F.~Heide, S.~Diamond, D.~B. Lindell, and G.~Wetzstein, ``Sub-picosecond photon-efficient 3d imaging using single-photon sensors,'' \emph{Scientific reports}, vol.~8, no.~1, p. 17726, 2018.

\bibitem{gyongy2020high}
I.~Gyongy, S.~W. Hutchings, A.~Halimi, M.~Tyler, S.~Chan, F.~Zhu, S.~McLaughlin, R.~K. Henderson, and J.~Leach, ``High-speed 3d sensing via hybrid-mode imaging and guided upsampling,'' \emph{Optica}, vol.~7, no.~10, pp. 1253--1260, 2020.

\bibitem{gariepy2016picosecond}
G.~Gariepy, J.~Leach, R.~Warburton, S.~Chan, R.~Henderson, and D.~Faccio, ``Picosecond time-resolved imaging using spad cameras,'' in \emph{Emerging Imaging and Sensing Technologies}, vol. 9992.\hskip 1em plus 0.5em minus 0.4em\relax SPIE, 2016, pp. 130--137.

\bibitem{gutierrez2023learned}
F.~Gutierrez-Barragan, F.~Mu, A.~Ardelean, A.~Ingle, C.~Bruschini, E.~Charbon, Y.~Li, M.~Gupta, and A.~Velten, ``Learned compressive representations for single-photon 3d imaging,'' in \emph{Proceedings of the IEEE/CVF International Conference on Computer Vision}, 2023, pp. 10\,756--10\,766.

\bibitem{ren2018high}
X.~Ren, P.~W. Connolly, A.~Halimi, Y.~Altmann, S.~McLaughlin, I.~Gyongy, R.~K. Henderson, and G.~S. Buller, ``High-resolution depth profiling using a range-gated cmos spad quanta image sensor,'' \emph{Optics express}, vol.~26, no.~5, pp. 5541--5557, 2018.

\bibitem{dutton2015spad}
N.~A. Dutton, I.~Gyongy, L.~Parmesan, S.~Gnecchi, N.~Calder, B.~R. Rae, S.~Pellegrini, L.~A. Grant, and R.~K. Henderson, ``A spad-based qvga image sensor for single-photon counting and quanta imaging,'' \emph{IEEE Transactions on Electron Devices}, vol.~63, no.~1, pp. 189--196, 2015.

\bibitem{erdogan2019cmos}
A.~T. Erdogan, R.~Walker, N.~Finlayson, N.~Krstaji{\'c}, G.~Williams, J.~Girkin, and R.~Henderson, ``A cmos spad line sensor with per-pixel histogramming tdc for time-resolved multispectral imaging,'' \emph{IEEE Journal of Solid-State Circuits}, vol.~54, no.~6, pp. 1705--1719, 2019.

\bibitem{dutton2016single}
N.~A. Dutton, I.~Gyongy, L.~Parmesan, and R.~K. Henderson, ``Single photon counting performance and noise analysis of cmos spad-based image sensors,'' \emph{Sensors}, vol.~16, no.~7, p. 1122, 2016.

\bibitem{zhang2021240}
C.~Zhang, N.~Zhang, Z.~Ma, L.~Wang, Y.~Qin, J.~Jia, and K.~Zang, ``A 240$\times$ 160 3d-stacked spad dtof image sensor with rolling shutter and in-pixel histogram for mobile devices,'' \emph{IEEE Open Journal of the Solid-State Circuits Society}, vol.~2, pp. 3--11, 2021.

\bibitem{taneski2023guided}
F.~Taneski, I.~Gyongy, T.~Al~Abbas, and R.~K. Henderson, ``Guided direct time-of-flight lidar using stereo cameras for enhanced laser power efficiency,'' \emph{Sensors}, vol.~23, no.~21, p. 8943, 2023.

\bibitem{ingle2023count}
A.~Ingle and D.~Maier, ``Count-free single-photon 3d imaging with race logic,'' \emph{IEEE Transactions on Pattern Analysis and Machine Intelligence}, 2023.

\bibitem{po2022adaptive}
R.~Po, A.~Pediredla, and I.~Gkioulekas, ``Adaptive gating for single-photon 3d imaging,'' in \emph{Proceedings of the IEEE/CVF Conference on Computer Vision and Pattern Recognition}, 2022, pp. 16\,354--16\,363.

\bibitem{battrawy2019lidar}
R.~Battrawy, R.~Schuster, O.~Wasenm{\"u}ller, Q.~Rao, and D.~Stricker, ``Lidar-flow: Dense scene flow estimation from sparse lidar and stereo images,'' in \emph{2019 IEEE/RSJ International Conference on Intelligent Robots and Systems (IROS)}.\hskip 1em plus 0.5em minus 0.4em\relax IEEE, 2019, pp. 7762--7769.

\bibitem{chen2018estimating}
Z.~Chen, V.~Badrinarayanan, G.~Drozdov, and A.~Rabinovich, ``Estimating depth from rgb and sparse sensing,'' in \emph{Proceedings of the European Conference on Computer Vision (ECCV)}, 2018, pp. 167--182.

\bibitem{mal2018sparse}
F.~Mal and S.~Karaman, ``Sparse-to-dense: Depth prediction from sparse depth samples and a single image,'' in \emph{2018 IEEE International Conference on Robotics and Automation (ICRA)}.\hskip 1em plus 0.5em minus 0.4em\relax IEEE, 2018, pp. 1--8.

\bibitem{lu2015sparse}
J.~Lu and D.~Forsyth, ``Sparse depth super resolution,'' in \emph{Proceedings of the IEEE Conference on Computer Vision and Pattern Recognition}, 2015, pp. 2245--2253.

\bibitem{uhrig2017sparsity}
J.~Uhrig, N.~Schneider, L.~Schneider, U.~Franke, T.~Brox, and A.~Geiger, ``Sparsity invariant cnns,'' in \emph{2017 International Conference on 3D Vision (3DV)}.\hskip 1em plus 0.5em minus 0.4em\relax IEEE, 2017, pp. 11--20.

\bibitem{riegler2016atgv}
G.~Riegler, M.~R{\"u}ther, and H.~Bischof, ``Atgv-net: Accurate depth super-resolution,'' in \emph{European Conference on Computer Vision}.\hskip 1em plus 0.5em minus 0.4em\relax Springer, 2016, pp. 268--284.

\bibitem{hui16msgnet}
T.-W. Hui, C.~C. Loy, and X.~Tang, ``Depth map super-resolution by deep multi-scale guidance,'' in \emph{Proceedings of European Conference on Computer Vision (ECCV)}, 2016.

\bibitem{van2019sparse}
W.~Van~Gansbeke, D.~Neven, B.~De~Brabandere, and L.~Van~Gool, ``Sparse and noisy lidar completion with rgb guidance and uncertainty,'' \emph{arXiv preprint arXiv:1902.05356}, 2019.

\bibitem{gruber2019gated2depth}
T.~Gruber, F.~Julca-Aguilar, M.~Bijelic, W.~Ritter, K.~Dietmayer, and F.~Heide, ``Gated2depth: Real-time dense lidar from gated images,'' \emph{arXiv preprint arXiv:1902.04997}, 2019.

\bibitem{tilmon2021saccadecam}
B.~Tilmon and S.~J. Koppal, ``Saccadecam: Adaptive visual attention for monocular depth sensing,'' in \emph{Proceedings of the IEEE/CVF International Conference on Computer Vision}, 2021, pp. 6009--6018.

\bibitem{yamamoto2018efficient}
T.~Yamamoto, Y.~Kawanishi, I.~Ide, H.~Murase, F.~Shinmura, and D.~Deguchi, ``Efficient pedestrian scanning by active scan lidar,'' in \emph{Advanced Image Technology (IWAIT), 2018 International Workshop on}.\hskip 1em plus 0.5em minus 0.4em\relax IEEE, 2018, pp. 1--4.

\bibitem{tasneem2018dirrectionally}
Z.~Tasneem, D.~Wang, H.~Xie, and S.~J. Koppal, ``Directionally controlled time-of-flight ranging for mobile sensing platforms,'' in \emph{Robotics: Science and Systems}, 2018.

\bibitem{adaptivelidarstanford}
A.~Bergman, D.~Lindell, and G.~Wetzstein, ``Deep adaptive lidar: End-to-end optimization of sampling and depth completion at low sampling rates,'' \emph{ICCP}, 2020.

\bibitem{pittaluga2020towards}
F.~Pittaluga, Z.~Tasneem, J.~Folden, B.~Tilmon, A.~Chakrabarti, and S.~J. Koppal, ``Towards a mems-based adaptive lidar,'' in \emph{2020 International Conference on 3D Vision (3DV)}.\hskip 1em plus 0.5em minus 0.4em\relax IEEE, 2020, pp. 1216--1226.

\bibitem{guenter2012foveated}
B.~Guenter, M.~Finch, S.~Drucker, D.~Tan, and J.~Snyder, ``Foveated 3d graphics,'' \emph{ACM transactions on Graphics (tOG)}, vol.~31, no.~6, pp. 1--10, 2012.

\bibitem{albert2017latency}
R.~Albert, A.~Patney, D.~Luebke, and J.~Kim, ``Latency requirements for foveated rendering in virtual reality,'' \emph{ACM Transactions on Applied Perception (TAP)}, vol.~14, no.~4, pp. 1--13, 2017.

\bibitem{tursun2019luminance}
O.~T. Tursun, E.~Arabadzhiyska-Koleva, M.~Wernikowski, R.~Mantiuk, H.-P. Seidel, K.~Myszkowski, and P.~Didyk, ``Luminance-contrast-aware foveated rendering,'' \emph{ACM Transactions on Graphics (TOG)}, vol.~38, no.~4, pp. 1--14, 2019.

\bibitem{huang2015light}
F.-C. Huang, D.~P. Luebke, and G.~Wetzstein, ``The light field stereoscope.'' in \emph{SIGGRAPH emerging technologies}, 2015, pp. 24--1.

\bibitem{sun2017perceptually}
Q.~Sun, F.-C. Huang, J.~Kim, L.-Y. Wei, D.~Luebke, and A.~Kaufman, ``Perceptually-guided foveation for light field displays,'' \emph{ACM Transactions on Graphics (TOG)}, vol.~36, no.~6, pp. 1--13, 2017.

\bibitem{patney2016perceptually}
A.~Patney, J.~Kim, M.~Salvi, A.~Kaplanyan, C.~Wyman, N.~Benty, A.~Lefohn, and D.~Luebke, ``Perceptually-based foveated virtual reality,'' in \emph{ACM SIGGRAPH 2016 emerging technologies}, 2016, pp. 1--2.

\bibitem{meng20203d}
X.~Meng, R.~Du, J.~F. JaJa, and A.~Varshney, ``3d-kernel foveated rendering for light fields,'' \emph{IEEE transactions on visualization and computer graphics}, vol.~27, no.~8, pp. 3350--3360, 2020.

\bibitem{sheehan2021sketching}
\BIBentryALTinterwordspacing
M.~Sheehan, J.~Tachella, and M.~Davies, ``A sketching framework for reduced data transfer in photon counting lidar,'' \emph{IEEE Transactions on Computational Imaging}, vol.~7, p. 989–1004, 2021. [Online]. Available: \url{http://dx.doi.org/10.1109/TCI.2021.3113495}
\BIBentrySTDinterwordspacing

\bibitem{zhang2022firstarrival}
T.~Zhang, M.~J. White, A.~Dave, S.~Ghajari, A.~Raghuram, A.~C. Molnar, and A.~Veeraraghavan, ``First arrival differential lidar,'' in \emph{2022 IEEE International Conference on Computational Photography (ICCP)}, 2022, pp. 1--12.

\bibitem{white2022differentialspad}
M.~White, S.~Ghajari, T.~Zhang, A.~Dave, A.~Veeraraghavan, and A.~Molnar, ``A differential spad array architecture in 0.18 um cmos for hdr imaging,'' in \emph{2022 IEEE International Symposium on Circuits and Systems (ISCAS)}, 2022, pp. 292--296.

\bibitem{totini2023histogramlesslidar}
A.~Tontini, S.~Mazzucchi, R.~Passerone, N.~Broseghini, and L.~Gasparini, ``Histogram-less lidar through spad response linearization,'' \emph{IEEE Sensors Journal}, vol.~PP, pp. 1--1, 01 2023.

\bibitem{sun2020superresspad}
\BIBentryALTinterwordspacing
Q.~Sun, J.~Zhang, X.~Dun, B.~Ghanem, Y.~Peng, and W.~Heidrich, ``End-to-end learned, optically coded super-resolution spad camera,'' \emph{ACM Trans. Graph.}, vol.~39, no.~2, mar 2020. [Online]. Available: \url{https://doi.org/10.1145/3372261}
\BIBentrySTDinterwordspacing

\bibitem{gupta2019asynchronous}
A.~Gupta, A.~Ingle, and M.~Gupta, ``Asynchronous single-photon 3d imaging,'' in \emph{Proceedings of the IEEE/CVF International Conference on Computer Vision}, 2019, pp. 7909--7918.

\bibitem{gupta2019photon}
A.~Gupta, A.~Ingle, A.~Velten, and M.~Gupta, ``Photon-flooded single-photon 3d cameras,'' in \emph{Proceedings of the IEEE/CVF Conference on Computer Vision and Pattern Recognition}, 2019, pp. 6770--6779.

\bibitem{gutierrez2021itof2dtof}
F.~Gutierrez-Barragan, H.~Chen, M.~Gupta, A.~Velten, and J.~Gu, ``itof2dtof: A robust and flexible representation for data-driven time-of-flight imaging,'' \emph{IEEE Transactions on Computational Imaging}, vol.~7, pp. 1205--1214, 2021.

\bibitem{SilbermanECCV12NYUV2}
P.~K. Nathan~Silberman, Derek~Hoiem and R.~Fergus, ``Indoor segmentation and support inference from rgbd images,'' in \emph{ECCV}, 2012.

\bibitem{Geiger2012CVPRKITTI}
A.~Geiger, P.~Lenz, and R.~Urtasun, ``Are we ready for autonomous driving? the kitti vision benchmark suite,'' in \emph{Conference on Computer Vision and Pattern Recognition (CVPR)}, 2012.

\bibitem{bhat23mono}
\BIBentryALTinterwordspacing
S.~F. Bhat, R.~Birkl, D.~Wofk, P.~Wonka, and M.~Müller, ``Zoedepth: Zero-shot transfer by combining relative and metric depth,'' 2023. [Online]. Available: \url{https://arxiv.org/abs/2302.12288}
\BIBentrySTDinterwordspacing

\bibitem{lee2023caspi}
J.~Lee, A.~Ingle, J.~V. Chacko, K.~W. Eliceiri, and M.~Gupta, ``Caspi: collaborative photon processing for active single-photon imaging,'' \emph{Nature Communications}, vol.~14, no.~1, p. 3158, 2023.

\bibitem{beer2018spad}
M.~Beer, O.~M. Schrey, J.~F. Haase, J.~Ruskowski, W.~Brockherde, B.~J. Hosticka, and R.~Kokozinski, ``Spad-based flash lidar sensor with high ambient light rejection for automotive applications,'' in \emph{Quantum Sensing and Nano Electronics and Photonics XV}, vol. 10540.\hskip 1em plus 0.5em minus 0.4em\relax SPIE, 2018, pp. 320--327.

\bibitem{dosovitskiy2017carla}
A.~Dosovitskiy, G.~Ros, F.~Codevilla, A.~Lopez, and V.~Koltun, ``Carla: An open urban driving simulator,'' in \emph{Conference on robot learning}.\hskip 1em plus 0.5em minus 0.4em\relax PMLR, 2017, pp. 1--16.

\bibitem{burri2017linospad}
S.~Burri, C.~Bruschini, and E.~Charbon, ``Linospad: a compact linear spad camera system with 64 fpga-based tdc modules for versatile 50 ps resolution time-resolved imaging,'' \emph{Instruments}, vol.~1, no.~1, p.~6, 2017.

\bibitem{snic_cvpr17}
R.~Achanta and S.~Susstrunk, ``Superpixels and polygons using simple non-iterative clustering,'' in \emph{IEEE Conference on Computer Vision and Pattern Recognition (CVPR)}, 2017.

\bibitem{ardelean2023computational}
A.~Ardelean, ``Computational imaging spad cameras,'' Ph.D. dissertation, EPFL, 2023.

\bibitem{sundar2023sodacam}
V.~Sundar, A.~Ardelean, T.~Swedish, C.~Bruschini, E.~Charbon, and M.~Gupta, ``Sodacam: Software-defined cameras via single-photon imaging,'' in \emph{Proceedings of the IEEE/CVF International Conference on Computer Vision}, 2023, pp. 8165--8176.

\bibitem{callenberg2021low}
C.~Callenberg, Z.~Shi, F.~Heide, and M.~B. Hullin, ``Low-cost spad sensing for non-line-of-sight tracking, material classification and depth imaging,'' \emph{ACM Transactions on Graphics (TOG)}, vol.~40, no.~4, pp. 1--12, 2021.

\end{thebibliography}

% \newpage
% \begin{IEEEbiography}{Michael Shell}
% Biography text here.
% \end{IEEEbiography}

% insert where needed to balance the two columns on the last page with
% biographies
%\newpage

% % if you will not have a photo at all:
% \begin{IEEEbiographynophoto}{John Doe}
% Biography text here.
% \end{IEEEbiographynophoto}

% You can push biographies down or up by placing
% a \vfill before or after them. The appropriate
% use of \vfill depends on what kind of text is
% on the last page and whether or not the columns
% are being equalized.
%\vfill

\end{document}

%%
%% The "title" command has an optional parameter,
%% allowing the author to define a "short title" to be used in page headers.
\title{FoveaSPAD: Exploiting Depth Priors for Adaptive
and Efficient Single-Photon 3D Imaging}
\maketitle

\section{Worst Case Stochastic Limits}
\label{sec:7worst}
In this section, we characterize the worst case scenario where depth is wrongly detected by a foveated SPAD pixels.
This lower bound helps us understand the limits of the approach.
However, it is different from a best or average case analysis, which would be useful for deployment, and we leave such analysis to future work. 

Foveation errors in our framework may be due to monocular depth calibration errors, ambient light, and global effects such as multi-bounce inter-reflections.
In these scenarios, the window predicted by foveation may not overlap with the expected transient peak.
To characterize these errors, we use an analysis method described in Gupta et al. to find the probability that a peak will be detected in the set of foveated bins. 

Consider the initial foveation window of $M$ bins for all the $S$ pixels in the camera.
We define $p_\text{gt}$ as the probability that a detected photon originated from the laser dot that illuminates the scene point of interest.
We also defined $p_\text{multipath}$ as the probability that a detected photon experienced multipath bounces and $p_\text{floor}$ as the probability that the sensor noise does not create spurious peaks.

First, we consider the probability that the direct, single bounce photon from the laser to the scene point was detected in the $M$ binned foveation window --- this is the definition of $p_\text{gt}$. 
We also consider photons from the laser that experience multipath effects, which we model as $p_\text{gt} p_\text{multipath}$.
The foveation window must have none of these multipath photons from any of the other $M-1$ bins that originate at the laser.
Further, the noise floor must be low for this detection. In other words, the probability of peak detection is 

\begin{equation}
   p_\text{gt} (1 - p_\text{gt} p_\text{multipath})^{M-1} p_\text{floor} . 
\end{equation} 

We now model the worst case scenario, where none of the $S$ pixels get the correct foveated depth. The chances that this happens are:

\begin{equation}
   p_\text{worst} = (1 - p_\text{gt} (1 - p_\text{gt} p_\text{multipath})^{M-1} p_\text{floor})^S. 
\end{equation}

As in Gupta et al. , we set $\frac{\delta p_\text{worst}}{\delta p_\text{gt}} = 0$ to analyze when this worst case probability is maximized. As we show in appendix A, this simplifies to the following:

\begin{equation}
\begin{split}
   & S (1 - p_\text{gt} (1 - p_\text{gt} p_\text{multipath})^{M-1} p_\text{floor})^{S-1} \  \cdot \\ & (-p_\text{floor}) ((1 - p_\text{gt} p_\text{multipath})^{M-2}  \  \cdot \\ & ((1 - M p_\text{gt} (p_\text{multipath}) ) 
    = 0
\end{split}
\end{equation}

%\begin{equation}
%   -p_{noise} ((1 - p_{gt} p_{multipath})^{M-1} + (M-1) p_{gt} (1 - p_{gt} p_{multipath})^{M-2} (-p_{multipath}) )
%\end{equation}

%\begin{equation}
 %  -p_{noise} (1 - p_{gt} p_{multipath})^{M-2}  ((1 - p_{gt} p_{multipath}) + (M-1) p_{gt} (-p_{multipath}) )
%\end{equation}

%\begin{equation}
%   -p_{noise} (1 - p_{gt} p_{multipath})^{M-2}  ((1 - M p_{gt} (p_{multipath}) )
%\end{equation}

We now explain how this relation can be used in practice. Recall that $p_\text{gt}$ is the probability that laser photons are detected, i.e. the chances that accurate depth recovery occurs. Only two values of $p_\text{gt}$ make the above worst case relation zero. The first term to zero out the relation is that $p_\text{gt} = \frac{1}{p_\text{multipath}}$. From the definition of probability, this is only possible if the probability is 1 for every bin to have both photons from the laser and have multipath effects --- i.e. the scene is degenerate, such as made entirely from mirror BRDFs. 

The second possibility happens when  $p_\text{gt} = \frac{1}{M \ p_\text{multipath}}$, where the number of bins $M$ and the probability of multipath effects $p_\text{multipath}$ vary under the condition that $0 \le p_\text{gt} \le 1$. This suggests that heuristics to avoid the worst case, where ideal bands of foveated windows $M$ can be used for scenes with particular global illumination characteristics denoted by $p_\text{multipath}$. 

As an example, consider a set of bins $M=1000$. Consider a situation where multipath effects are very low, and $p_\text{multipath} = 0.001$. In this scenario, the probability of accurate depth recovery is $p_\text{gt} = 1$, which is the case in our simulated results where there are no multi-path effects. However, if the multipath effects are, say one in ten, then $p_\text{gt} = 0.1$ then the probability of depth recovery falls, in the worst case to $p_\text{gt} = 0.01$. Attempting to improve the probability of detecting laser photons $p_\text{gt}$ by varying the number of bins cannot be done without reducing depth resolution.

In Fig. \ref{fig:fig_sim}(a) we show a verification of Eq 4 from the main text.
We varied the exposure, foveation interval $M$, and computed SSD for one scene. 
These simulations show that depth quality does not increase linearly with increase in foveation bins, but does so with exposure, as predicted by Eq 4.
In \ref{fig:fig_sim}(b), we have also shown verification for Eq. 10 for the degenerate mode in black ($p_\text{worst}$ = 1) and for the recommended mode $p_\text{gt} = \frac{1}{M \ p_\text{multipath}}$.
$p_\text{gt}$ and $p_\text{floor}$ were modeled as Gaussians and $p_\text{multipath}$ is shown for two cases, high (green) and low (red).
As the graph shows, with lower probabilities of $p_\text{multipath}$, tighter foveation intervals are possible even in these upper bounds of worst cases.

Given this worst-case analysis, we now tackle data from real SPAD sensors and scenes, which have noisy histogram floors, inter-reflections and other complex effects.  
\begin{figure*}[!t]
\centering
\subfloat[]{\resizebox{0.4\textwidth}{!}{\input{Figs/RebuttalGraphs/depthFovea_SSD_Try.pgf}}%
\label{Eq10}} 
\hfil
\subfloat[]{\resizebox{0.4\textwidth}{!}{\input{Figs/RebuttalGraphs/EQ10_HighNLow.pgf}}%
\label{Eq4}}
\caption{\textbf{Eq. 4 and 10 validation:} (a) Depth foveation reduces bin width, reducing SNR.
 Increasing exposure can compensate for this SNR decrease (and improve the sum-squared difference SSD). (b) The red and green curves show the upper bound on $p_\text{worst}$ from Eq.~10. These are generated based on nominal and worst case distributions of $p_\text{multipath}$, with $p_\text{gt} = \frac{1}{M \ p_\text{multipath}}. $}
\label{fig:fig_sim}
\end{figure*}

%Atul: not sure what the goal of this analysis is... are we trying to figure out the best M to choose when foveating? It seems to me that the practical utility of this result might be limited (likely because we considered the absolute worst case where none of the S pixels foveate correctly). Observe that p_gt = 1/Mp_multi. I posit that p_multi is going to be extremely small, say, 0.1%. Then for this to be a valid probability, M will have to be at least 1/0.1% = 1000, which is the span of the full histogram.

\section{Worst-Case Analysis}

We set $\frac{\delta p_\text{worst}}{\delta p_\text{gt}} = 0$ to analyze when this worst case probability is maximized:

\begin{equation}
\begin{split}
   & S (1 - p_\text{gt} (1 - p_\text{gt} p_\text{multipath})^{M-1} p_\text{noise})^{S-1} \  \cdot \\ & \frac{\delta}{\delta p_\text{gt}}(1 - p_\text{gt} (1 - p_\text{gt} p_\text{multipath})^{M-1} p_\text{noise}) = 0
\end{split}
\end{equation}

%\begin{equation}
%   S (1 - p_\text{gt} (1 - p_\text{gt} p_\text{multipath})^{M-1} p_\text{noise})^{S-1} \frac{\delta}{\delta p_\text{gt}}(1 - p_\text{gt} (1 - p_\text{gt} p_\text{multipath})^{M-1} p_\text{noise})
%\end{equation}

\begin{equation}
\begin{split}
   & S (1 - p_\text{gt} (1 - p_\text{gt} p_\text{multipath})^{M-1} p_\text{noise})^{S-1} \  \cdot \\ & (0 - \frac{\delta}{\delta p_\text{gt}}(p_\text{gt} (1 - p_\text{gt} p_\text{multipath})^{M-1} p_\text{noise}) = 0
\end{split}
\end{equation}

%\begin{equation}
%   (...) (0 - \frac{\delta}{\delta p_\text{gt}}(p_\text{gt} (1 - p_\text{gt} p_\text{multipath})^{M-1} p_\text{noise})
%\end{equation}

\begin{equation}
\begin{split}
   & S (1 - p_\text{gt} (1 - p_\text{gt} p_\text{multipath})^{M-1} p_\text{noise})^{S-1} (-p_\text{noise}) \  \cdot \\ & (1. (1 - p_\text{gt} p_\text{multipath})^{M-1} + p_\text{gt} \frac{\delta}{\delta p_\text{gt}}(1 - p_\text{gt} p_\text{multipath})^{M-1}) = 0
\end{split}
\end{equation}

%\begin{equation}
%   -p_\text{noise} (1. (1 - p_\text{gt} p_\text{multipath})^{M-1} + p_\text{gt} \frac{\delta}{\delta p_\text{gt}}(1 - p_\text{gt} p_\text{multipath})^{M-1})
%\end{equation}

\begin{equation}
\begin{split}
   & S (1 - p_\text{gt} (1 - p_\text{gt} p_\text{multipath})^{M-1} p_\text{noise})^{S-1} (-p_\text{noise}) \  \cdot \\ & ((1 - p_\text{gt} p_\text{multipath})^{M-1} + (M-1) p_\text{gt} \cdot \\ &  (1 - p_\text{gt} p_\text{multipath})^{M-2} \frac{\delta}{\delta p_\text{gt}} (1 - p_\text{gt} p_\text{multipath})) = 0
\end{split}
\end{equation}

%\begin{equation}
%   -p_\text{noise} ((1 - p_\text{gt} p_\text{multipath})^{M-1} + (M-1) p_\text{gt} (1 - p_\text{gt} p_\text{multipath})^{M-2} \frac{\delta}{\delta p_\text{gt}} (1 - p_\text{gt} p_\text{multipath})   )
%\end{equation}

\begin{equation}
\begin{split}
   & S (1 - p_\text{gt} (1 - p_\text{gt} p_\text{multipath})^{M-1} p_\text{noise})^{S-1} (-p_\text{noise}) \  \cdot \\ & ((1 - p_\text{gt} p_\text{multipath})^{M-1} + (M-1) p_\text{gt} \cdot \\ &  (1 - p_\text{gt} p_\text{multipath})^{M-2} (-p_\text{multipath}) ) = 0
\end{split}
\end{equation}

\begin{equation}
\begin{split}
   & S (1 - p_\text{gt} (1 - p_\text{gt} p_\text{multipath})^{M-1} p_\text{noise})^{S-1} \  \cdot \\ & (-p_\text{noise}) ((1 - p_\text{gt} p_\text{multipath})^{M-2}  \  \cdot \\ & ((1 - p_\text{gt} p_\text{multipath}) + (M-1) p_\text{gt} (-p_\text{multipath}) ) 
    = 0
\end{split}
\end{equation}

\begin{equation}
\begin{split}
   & S (1 - p_\text{gt} (1 - p_\text{gt} p_\text{multipath})^{M-1} p_\text{noise})^{S-1} \  \cdot \\ & (-p_\text{noise}) ((1 - p_\text{gt} p_\text{multipath})^{M-2}  \  \cdot \\ & ((1 - M p_\text{gt} (p_\text{multipath}) ) 
    = 0
\end{split}
\end{equation}

% \vspace{4cm}
\section{Memory Usage}

The memory usage experiments in the Suppl. Table \ref{tab:memory_usage} and in all experiments throughout main text, full resolution indicates that the total number of bins $M = 1000$. Memory calculations were with simulation results using the NYUv2 dataset, the histogram images at full resolution being of size $(640,480,1000)$.

Suppl. Table \ref{tab:memory_usage_second} shows the memory usage for the spatio-temporal algorithm in experiments varying the number of sampled pixels.
\begin{table}[ht]
\centering
\caption{\textbf{Memory Usage:} Memory Foveation experiments.}
\label{tab:memory_usage}\begin{tabular}{cc}
\hline
\textbf{Histogram Resolution} & \textbf{Memory (MB)}\\
\hline
Full& 2343.75\\
1/4& 585.94\\
1/8& 292.97\\
1/16& 145.31\\

\hline
\end{tabular}

\end{table}

\begin{table}[ht]
\centering
\caption{\textbf{Memory Usage:} Spatio-Temporal experiments at 1/16 M }
\label{tab:memory_usage_second}\begin{tabular}{cccc}
\hline
\textbf{Num. Pixels} & \textbf{Num. Buckets} & \textbf{Memory (MB)} & \textbf{\% of Total Pix.} \\
\hline
50&32& 0.76 & 0.52\% \\
100&32& 1.51 & 1.04\% \\
500&32 & 7.57 & 5.21\% \\
50&64 & 1.51 & 1.04\% \\
100&64 & 3.03 & 2.08\% \\
500&64 & 15.14 & 10.42\% \\

\hline
\end{tabular}

\end{table}

% \vspace{cm}
\section{Error Masks for Memory Foveation}
In Suppl. Fig.~\ref{fig:errormasks} we show example error masks for two different scenes from NYUv2 dataset shown in Fig.~2 in the main text. We spatially downscale the true depth maps by a factor of 4$\times$ and compute depth errors in our memory-foveated results. Observe that this reveals some regions around boundaries and object edges where the error is large. This information can be used in a feedback loop by changing the foveation strategy and drive the error down. For example, we can use a wider foveation window.
\begin{figure}[H]
\centering
 \includegraphics[width=\columnwidth]{Figs/errormasks.png}
 \caption{\textbf{Error Masks.} The absolute distance errors for two scenes from the NYUv2 dataset show depth errors around object edges. Brighter pixels show higher absolute error for memory foveation.}
 %\Description{Our technique for saving memory, and improving depth resolution for SPAD cameras.}
 \label{fig:errormasks}
\end{figure}